\pdfoutput=1

\documentclass[%
 aip,
 amsmath,amssymb,
 onecolumn,
 preprint,
 superscriptaddress,
]{revtex4-1}

\usepackage{graphicx}
\usepackage{dcolumn}
\usepackage{bm}
\usepackage[utf8]{inputenc}
\usepackage[T1]{fontenc}
\usepackage{mathptmx}
\usepackage{etoolbox}

\makeatletter
\def\@email#1#2{%
 \endgroup
 \patchcmd{\titleblock@produce}
  {\frontmatter@RRAPformat}
  {\frontmatter@RRAPformat{\produce@RRAP{*#1\href{mailto:#2}{#2}}}\frontmatter@RRAPformat}
  {}{}
}%
\makeatother

\begin{document}

\title[]{Reciprocal Space Approach to Dipolarly Coupled Magnetic Hetero-Structures}


\author{\underline{A. Del Giacco}}
\affiliation{Department of Materials Science and Engineering, Massachusetts Institute of Technology, Cambridge, Massachusetts 02139, USA}
\affiliation{Department of Physics, Politecnico di MilanoPiazza Leonardo da Vinci 32, Milan 20133, Italy}
\affiliation{Institute for NanoSystems Innovation (NanoSI), Northeastern University, Boston, MA 02115, USA}
\email{andreadg@mit.edu}

\author{\underline{L. Menna}}
\affiliation{Department of Physics, Politecnico di MilanoPiazza Leonardo da Vinci 32, Milan 20133, Italy}
\affiliation{Institute of Photonics and Quantum Sciences, SUPA, Heriot-Watt University, Edinburgh EH14 4AS, United Kingdom}

\author{M. J. Gross}
\affiliation{Department of Electrical Engineering and Computer Science, Massachusetts Institute of Technology, Cambridge, Massachusetts 02139, USA}
\affiliation{Department of Materials Science and Engineering, Massachusetts Institute of Technology, Cambridge, Massachusetts 02139, USA}

\author{O. Wojewoda}
\affiliation{Department of Materials Science and Engineering, Massachusetts Institute of Technology, Cambridge, Massachusetts 02139, USA}
\affiliation{CEITEC BUT, Brno University of Technology, Brno, Czech Republic}

\author{V. Levati}
\affiliation{Department of Physics, Politecnico di MilanoPiazza Leonardo da Vinci 32, Milan 20133, Italy}

\author{M. Urbánek}
\affiliation{CEITEC BUT, Brno University of Technology, Brno, Czech Republic}

\author{S. Kurdi}
\affiliation{Institute of Photonics and Quantum Sciences, SUPA, Heriot-Watt University, Edinburgh EH14 4AS, United Kingdom}

\author{E. Albisetti}
\affiliation{Department of Physics, Politecnico di MilanoPiazza Leonardo da Vinci 32, Milan 20133, Italy}

\author{D. Petti}
\affiliation{Department of Physics, Politecnico di MilanoPiazza Leonardo da Vinci 32, Milan 20133, Italy}

\author{C. A. Ross}

\affiliation{Department of Materials Science and Engineering, Massachusetts Institute of Technology, Cambridge, Massachusetts 02139, USA}
\begin{abstract}
We present an analytical framework capable of describing spin-waves dynamic in magnetic hetero-structures composed of a pair of exchange-decoupled magnetic layers separated by a nonmagnetic spacer, focusing in particular on garnet-based multilayers. The model captures the formation of collective spin-wave modes, namely symmetric and antisymmetric,  arising from dipolar coupling and provides direct access to the dispersion relation of the system and consequent interference phenomena. This formalism establishes a versatile theoretical tool for the predictive design of dipolarly coupled magnonic devices, providing access to their eigenfrequencies and mode shapes.
\end{abstract}

\maketitle

\section{\label{sec:Intro}Introduction\protect\\}
The design of magnetic hetero-structures lies at the core of spintronics and nanomagnetism. Central to this effort is the ability to engineer spin transport through careful control of interfacial properties, material selection, and layer composition \cite{GarnettSuperLattice,fourmont2025exchangeFerroAntiFErroHeter,gross2025interfacial,MagnetHeavyMetal,ANIS-fan2025dynamically3Layer,wu2016observationYIGPt,pirro2014spin_YIGPt,klingler2018spinYIGCO,qian2024unraveling_YIGPy}. Such hetero-structures represent a versatile platform for engineering the magnonic dispersion relation and introduce tailored magnetic anisotropies capable of breaking spin-wave reciprocity, enabling compact and reconfigurable magnonic devices that offer directional control, signal routing, and interference-based functionalities \cite{Nonlocalmagnetizationdynamicstserkovnyak2005nonlocal,althammerYIGPt,SpinHallmagnetoresistance-nakayama2013spin,milloSWSAF1,SAF3-gallardo2019reconfigurable,SAF4-gallardo2021spin,chen2013theorySpinHallMagnetoR,gladiiSAF2,kruglyak2010magnonics,Chumak2015-so}.Their theoretical description builds upon foundational studies of single-layer thin-film magnets\cite{KalinikosSlavin,THEORY-SINGLE_HEIS} and it has been progressively extended to multilayered systems \cite{THEORY-MULTI_FILMS_OLD,THEORY-MULTI_FILMS,THEORY_MULTI_INTER_EXC,THEORY-BILAYER_EXC_COUPLING,THEORY-MULTI_HILLEBRANDS,BILAYER_GALLARDO}. 

In this work, we focus on magnetic hetero-structures composed of two exchange-decoupled magnetic layers separated by a non magnetic spacer, describing their collective behavior and its dependence on the dipolar coupling.The nontrivial form of the interaction  leads to the formation of collective spin-wave modes. Optical and acoustic branches naturally emerge in the dispersion modes, accompanied by range of  effects including interference phenomena \cite{MS-qin_nanoscale_2021,doi:10.1126/sciadv.1701517}, and tunable energy transfer.
Consistent with previous studies on lateral dipolarly coupled waveguides \cite{doi:10.1126/sciadv.1701517} and with vertical backward-volume configurations in permalloy hetero-structures \cite{szulc_magnetic_2025}, vertically stacked magnetic systems support the simultaneous excitation of spin-wave eigenmodes characterized by different wavevectors $k_s$ and $k_{as}$, corresponding to collective symmetric and antisymmetric oscillations. Their resulting wavevector separation $\Delta k_x = k_{s} - k_{as}$ leads to interference and a transfer of energy between the two magnetic layers over a coupling length $L_{\mathrm{coupling}} \approx \pi / |\Delta k_x|$. The resulting spatial modulation of the averaged spin-wave intensity arises from this coupled dynamics.

Vertical coupling breaks symmetry along the out-of-plane direction. However, this problem cannot be treated as a simple lifting of the degeneracy of the single-layer modes. In fact, such an approach fails to capture the spatially coupled nature of the dipolar interaction and its role in determining the symmetry and dispersion of the hetero-structure eigenmodes. Developing a compact analytical framework for dipolar-coupled systems is therefore essential for identifying the physical mechanisms governing the formation of collective spin-wave modes.

In contrast to laterally coupled waveguides, vertically coupled magnetic/non-magnetic/magnetic  (M/NM/M) waveguides can exhibit a mode-degeneracy point at which the symmetric and antisymmetric branches intersect at $k\neq 0$. At the degeneracy point, the coupling length formally diverges, signaling a transition to an effectively uncoupled regime. This leads to coupling lengths that can vary potentially from infinite distances (near the degeneracy) to sub-$\mu$m, thus paving the way to a rich variety of tunable-purpose devices, such as power splitters and 3D power transfer architectures. \cite{szulc_magnetic_2025,Gubbiotti_2025}. 
\begin{figure}
    \centering
    \includegraphics[width=0.5\linewidth]{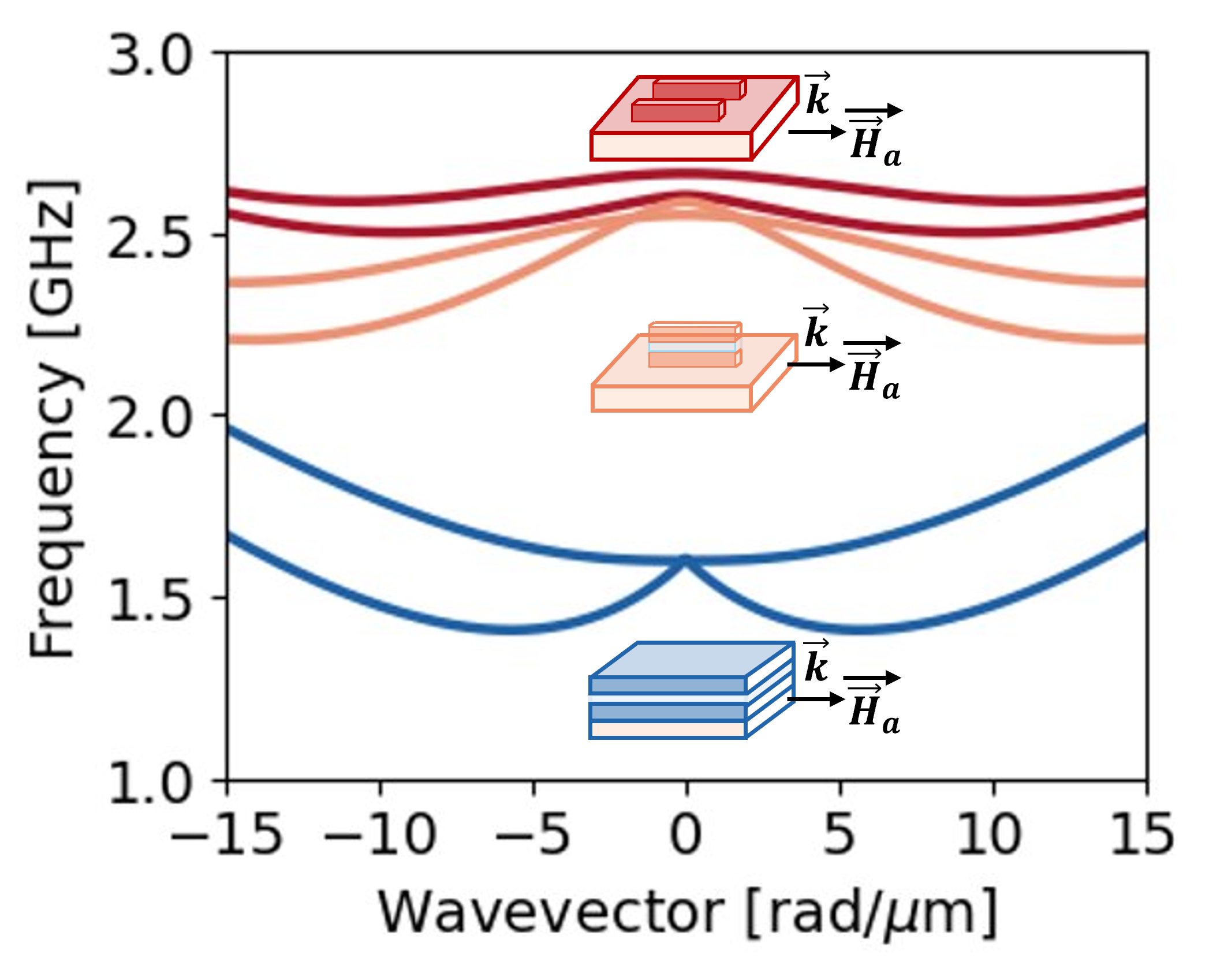}
    \caption{\label{fig:LAT_VERT} Comparison of the the spin-wave dispersion relation between laterally coupled waveguides (red), vertically coupled hetero-structure waveguides (orange), and an infinite film  hetero-structure (blue), in backward volume configuration. The vertical hetero-structure (blue and orange) is calculated for YIG(100 nm)/NM(50 nm)/YIG(100 nm) in which the vertical waveguides has width $w=100$ nm. The lateral waveguide configuration (red) has two waveguides of thickness $t= 50$ nm of YIG separated by $50$  nm.}
\end{figure}

In this work, we introduce an analytical model for spin waves in vertically coupled magnetic layers separated by a non-magnetic spacer that suppresses direct exchange interaction. The model exploits a Fourier-space approach, starting from the linearized Landau-Lifshitz equation, and extends formalisms previously developed for planarly coupled waveguides\cite{VERBA_NANODOTS,doi:10.1126/sciadv.1701517}. By expressing the problem in terms of a tensorial operator that incorporates exchange, anisotropy, and both self and mutual dipolar interactions, we obtain a compact eigenvalue formulation that naturally yields symmetric and antisymmetric collective modes without invoking ad hoc splitting assumptions. The model is general to arbitrary relative orientations between the spin-wave wavevector and the equilibrium magnetization direction. We apply the formalism to recently investigated vertically stacked hetero-structures and benchmark the analytical results against micromagnetic simulations.

\section{\label{sec:Fourier_Space_Approach}Fourier Space Approach }

\subsection{\label{subsec:Hypothesis}Hypotheses and validity of the model}

The model relies on four main approximations. First, we consider hetero-structures that are uniformly magnetized, thereby excluding complex micromagnetic configurations such as multidomain states or noncollinear textures. Second, the magnetization profile along the out-of-plane direction is assumed to be uniform across the thickness of each layer, restricting the analysis to the fundamental volume mode ($n = 0$). Third, the spin-wave modes are assumed to be symmetric across the width of the waveguides \cite{doi:10.1126/sciadv.1701517}. Finally, magnetic damping is not included in the present discussion.

These four assumptions enable an analytical evaluation of the demagnetizing tensor in Fourier space and lead to a compact eigenvalue formulation for the coupled spin-wave modes. As a consequence, the model does not capture phenomena associated with nonuniform thickness profiles, boundary pinning effects, or partially saturated magnetic states. In particular, deviations from full agreement with micromagnetic simulations are expected at increasing wavevectors, where thickness-mode hybridization and nonuniform magnetization profiles become progressively more relevant.

\subsection{\label{subsec:COUPLED_GENERAL}Dipolarly coupled waveguides}

The magnetization vector is defined within the small-signal approximation as
\begin{equation}
    \vec{M}(\vec{r},t)=M_s\vec{\mu}+\vec{m}(\vec{r},t),\quad \vec{m}(\vec{r},t)<<M_{s}\vec{\mu},
    \label{eq:M_linear}
\end{equation}
where $\vec{\mu}$ is the unit vector along the equilibrium magnetization direction, $M_s$ the saturation magnetization, and
$\vec{m}(\vec{r},t)=(m_x,m_y,m_z)$ is the space- and time-dependent dynamic component describing the magnetization precession in the linear regime.

By substituting Eq.~\eqref{eq:M_linear} into the Landau--Lifshitz equation and linearizing the resulting expression (see Supplementary S1), a Fourier-space representation\cite{KalinikosSlavin} for spin waves propagating along the $x$ direction is obtained,
\begin{equation}
-\mathrm{i}\,\omega_{k_x}^{(p)}\hat{\mathbf{I}}\vec{m}_{k_x}^{(p)}
=
\vec{\mu}\times
\sum_{q}\hat{\boldsymbol{\Omega}}_{k_x}^{(pq)}
\vec{m}_{k_x}^{(q)},
\label{eq:LL_linear}
\end{equation}
where $p,q\in\{1,2\}$ label the magnetic layers, $\hat{\mathbf{I}}$ is the identity matrix, $k_x$ the wave vector along the propagation direction, and $\hat{\boldsymbol{\Omega}}_{k_x}^{(pq)}$ is a tensor operator incorporating exchange, dipolar, and uniaxial anisotropy contributions. Solving this equation allows us to write the associated eigenvalue problem of the system in the frequency domain.

The general form of the operator $\hat{\boldsymbol{\Omega}}_{k_x}^{(pq)}$ reads
\begin{align}
\hat{\boldsymbol{\Omega}}_{k_x}^{(pq)}
&=
\gamma
\left[
B^{(p)}
+
\mu_0 M_s^{(p)}\left(\lambda_{\mathrm{ex}}^{(p)}\right)^2
\left(k_x^2+\kappa^2\right)
\right]
\delta_{pq}\,\hat{\mathbf{I}}
\nonumber\\
&\quad
-
\omega_M^{(p)}
\frac{2K^{(p)}}{\mu_0\left(M_s^{(p)}\right)^2}
\left(
\vec{u}^{(p)}\otimes\vec{u}^{(p)}
\right)
\delta_{pq}
+
\omega_M^{(q)}
\hat{\mathbf{F}}_{k_x}^{(pq)}(d_{pq}),
\label{eq:Omega_tensor}
\end{align}
where $\gamma$ is the gyromagnetic ratio and $\mu_0$ is the vacuum permeability. The quantity $B^{(p)}$ denotes the static internal magnetic field of layer $p$, while the exchange length is defined as
$\lambda_{\mathrm{ex}}=\sqrt{2A/(\mu_0 M_s^2)}$. Here $K^{(p)}$ and $\vec{u}^{(p)}$ are the anisotropy constant and anisotropy axis of layer $p$, respectively, and $d_{pq}$ is the distance between the centers of the magnetic layers along the thickness direction. The contribution of the transverse wavevector is accounted for by $\kappa=\pi/w$, while $\omega_M=\gamma\mu_0 M_s$ is the zero-field Larmor frequency.  The internal field $B^{(p)}$ \cite{https://doi.org/10.1002/aelm.202500575} can be written explicitly as 
\begin{align}
B^{(p)}
&=
\vec{B}_a\cdot\vec{\mu}
-
\mu_0 M_s^{(p)}
\vec{\mu}\cdot
\left[
\hat{\mathbf{F}}_{k_x=0}^{(pp)}(0)
+
\frac{M_s^{(q)}}{M_s^{(p)}}\hat{\mathbf{F}}_{k_x=0}^{(pq)}(d_{pq})
\right]
\vec{\mu}
\nonumber\\
&\quad
+
\vec{\mu}\cdot
\frac{2K^{(p)}}{M_s^{(p)}}
\left(
\vec{u}^{(p)}\otimes\vec{u}^{(p)}
\right)
\vec{\mu},
\label{eq:B_internal}
\end{align}
where $\vec{B}_{a}$ denotes the external magnetic field, $\hat{\mathbf{F}}_{k_x=0}^{(pp)}$ is the static self-demagnetizing tensor of the layer $p$, and $\hat{\mathbf{F}}_{k_x=0}^{(pq)}$ accounts for the correction arising from the embedding layer $p$ within the hetero-structure.
The tensor $\hat{\mathbf{F}}_{k_x}^{(pq)}$ describes the dipolar interaction in reciprocal space, including self- and mutual coupling between the layers.
Following Whang \textit{et al.} \cite{doi:10.1126/sciadv.1701517}, we exploit the Fourier-space approach developed by Beleggia \textit{et al.} \cite{BELEGGIA2004270} to compute the dipolar field components of the coupled system. The dipolar tensor elements are given by
\begin{widetext}
\begin{equation}
    \left(\hat{\mathbf{F}}_{k_x}^{(pq)}\right)_{ij}
    =
    \frac{1}{2\pi}
    \int_{-\infty}^{+\infty}
    \left(\hat{\mathbf{N}}_{k}^{(pq)}\right)_{ij}
    \mathrm{d}k_y,
    \quad
    i,j=x,y,z,
\end{equation}
\begin{equation}
    \label{eq:N}
    \hat{\mathbf{N}}_{k}^{(pq)}
    =
    \frac{|\sigma_{k}|^{2}}{\tilde{w}}
    \begin{pmatrix}
        \frac{k_{x}^{2}}{k^{2}}\Delta_{pq} &
        \frac{k_{x}k_{y}}{k^{2}}\Delta_{pq} &
        \mathrm{i}\frac{k_{x}}{k}\tilde{\Delta}_{pq}\,\mathrm{sign}(z_{p}-z_{q}) \\
        \frac{k_{x}k_{y}}{k^{2}}\Delta_{pq} &
        \frac{k_{y}^{2}}{k^{2}}\Delta_{pq} &
        i\frac{k_{y}}{k}\tilde{\Delta}_{pq}\mathrm{sign}(z_{p}-z_{q}) \\
        \mathrm{i}\frac{k_{x}}{k}\tilde{\Delta}_{pq}\,\mathrm{sign}(z_{p}-z_{q}) &
        i\frac{k_{y}}{k}\tilde{\Delta}_{pq}\mathrm{sign}(z_{p}-z_{q})&
        \delta_{pq}-\Delta_{pq}
    \end{pmatrix},
\end{equation}
\end{widetext}
with auxiliary functions,
\begin{gather*}
    \Delta_{pq}=\tilde{f}(kt_{p})+\left[f(kt_{p})-\tilde{f}(kt_{p})\right]\delta_{pq},
    \,
    \tilde{\Delta}_{pq}=\tilde{f}(kt_{p})\left(1-\delta_{pq}\right),
    \\
    f(kt_{p})=1-\frac{1-\mathrm{e}^{-kt_{p}}}{kt_{p}},
    \,
    \tilde{f}(kt_{p})=\frac{\left(1-\mathrm{e}^{-kt_{p}}\right)\left(1-\mathrm{e}^{-kt_{q}}\right)}{2kt_{p}}\mathrm{e}^{-k\delta},
\end{gather*}
where $k^{2}=k_{x}^{2}+k_{y}^{2}$, $\delta_{pq}$ is the Kroneker delta, $t_{q}$ indicates the thickness of layer $q$, $z_p$ and $z_q$ denote the $z$ positions of the centers of the two magnetic layers, $\sigma_k$ is the Fourier transform of the transverse spin-wave profile, and $\tilde{w}$ is the normalization constant associated with the transverse mode $m(y)$ \cite{doi:10.1126/sciadv.1701517}, and $\hat{\mathbf{N}}_{k}^{(pq)}$ the magnetometric tensor \cite{VERBA_NANODOTS} (the main steps of the derivation of the magnetometric tensor are reported in Supplementary S2).

In order to solve equation \ref{eq:LL_linear}, it is possible to note that  $\vec{\mu}\times\vec{v}$ (where $\vec{v}$ is a generic vector here) can be expressed in matrix form as:

\begin{equation}
    \vec{\mu}\times\vec{v}=\hat{\boldsymbol{\mu}}\vec{v},
    \quad
    \hat{\boldsymbol{\mu}}=
    \begin{pmatrix}
        0 & -\mu_{z} & \mu_{y}\\
        \mu_{z} & 0 & -\mu_{x} \\
        -\mu_{y} & \mu_{x} & 0
    \end{pmatrix}.
\end{equation}
Equation \ref{eq:LL_linear} can thus be written for layers $p=1,2$ expressed as:
\begin{equation}
    \begin{cases}
        -i\omega_{k_{x}}^{(1)}\hat{\mathbf{I}}\vec{m}_{k_{x}}^{(1)}=\hat{\mu}\left(\hat{\Omega}_{k_{x}}^{(11)}\vec{m}_{k_{x}}^{(1)}+\hat{\Omega}_{k_{x}}^{(12)}\vec{m}_{k_{x}}^{(2)}\right)\\
        -i\omega_{k_{x}}^{(2)}\hat{\mathbf{I}}\vec{m}_{k_{x}}^{(2)}=\hat{\mu}\left(\hat{\Omega}_{k_{x}}^{(21)}\vec{m}_{k_{x}}^{(1)}+\hat{\Omega}_{k_{x}}^{(22)}\vec{m}_{k_{x}}^{(2)}\right)\\
        
    \end{cases}
\end{equation}
with $\vec{m}_{k_{x}}=(m_x,m_y,m_z) \in \mathbb{R}^3$.
This allows us to reformulate the mathematical equations framework in an explicit eigenvalue problem:
\begin{equation}
-i\omega\,\hat{\mathbf{m}}= \hat{\boldsymbol{\mu}}_C \,\hat{\Omega}_C\,\hat{\mathbf{m}},
\end{equation}
with $\mathbf{m}_{k_x}\in\mathbb{R}^6$  referring to the three spatial components of the magnetization in  layers $p$ and $q$, while $\hat{\boldsymbol{\mu_C}}$ denotes the skew-symmetric matrix unveiling the coupling mechanism. This can be directly appreciated from the explicit tensor formalism below:
\begin{equation}
 -i\begin{pmatrix}
     \hat{\boldsymbol{\omega}} & 0 \\
     0 & \hat{\boldsymbol{\omega}}
 \end{pmatrix}
 \begin{pmatrix}
     \vec{m}^{(1)}\\
     \vec{m}^{(2)}
 \end{pmatrix}
 \begin{pmatrix}
     \hat{\boldsymbol{\mu}} & 0 \\
     0 & \hat{\boldsymbol{\mu}}
 \end{pmatrix}
 \begin{pmatrix}
     \hat{\boldsymbol{\Omega}}^{(11)} & \hat{\boldsymbol{\Omega}}^{(12)} \\
     \hat{\boldsymbol{\Omega}}^{(21)} & \hat{\boldsymbol{\Omega}}^{(22)}
 \end{pmatrix}
 \begin{pmatrix}
     \vec{m}^{(1)} \\
     \vec{m}^{(2)}
 \end{pmatrix}
\label{eq:TensorForm}
\end{equation}
with $\hat{\boldsymbol{\omega}}=\omega\hat{\mathbf{I}}_3$, and indices $k_{x}$ have been neglected to simplify the notation. 

This formulation enables a direct evaluation of the eigenmodes and dispersion relations in vertically coupled magnetic systems, where symmetric and antisymmetric spin-wave modes emerge from the dynamic dipolar interaction encoded in the off-diagonal tensors $\hat{\boldsymbol{\Omega}}^{(12)}$ and $\hat{\boldsymbol{\Omega}}^{(21)}$.

Within this framework, the associated eigenvalue problem can be solved, allowing direct access to the spin-wave dispersion relations and the corresponding dynamic magnetization profiles.
 Expressing the equilibrium magnetization direction in spherical coordinates as $\vec{\mu}=(\sin\theta\cos\phi,\sin\theta\sin\phi,\cos\theta)$ enables a unified treatment of arbitrary orientations of the applied magnetic field with respect to the propagation direction $\hat{x}$, under the full-saturation approximation. The substitution of $\vec{\mu}$ into the skew-symmetric matrix $\hat{\boldsymbol{\mu}}$ in Eq.~\eqref{eq:TensorForm} then allows the direct numerical evaluation of the eigenvalue problem.

Within this formalism, the model can be generalized to account for arbitrary relative orientations between the wavevector and the external magnetic field, as well as different layer thicknesses, saturation magnetizations, exchange constants, and anisotropy strengths and directions. The resulting eigenvalue problem can be efficiently solved using standard numerical approaches.
\begin{figure*}
\includegraphics[scale=0.5]{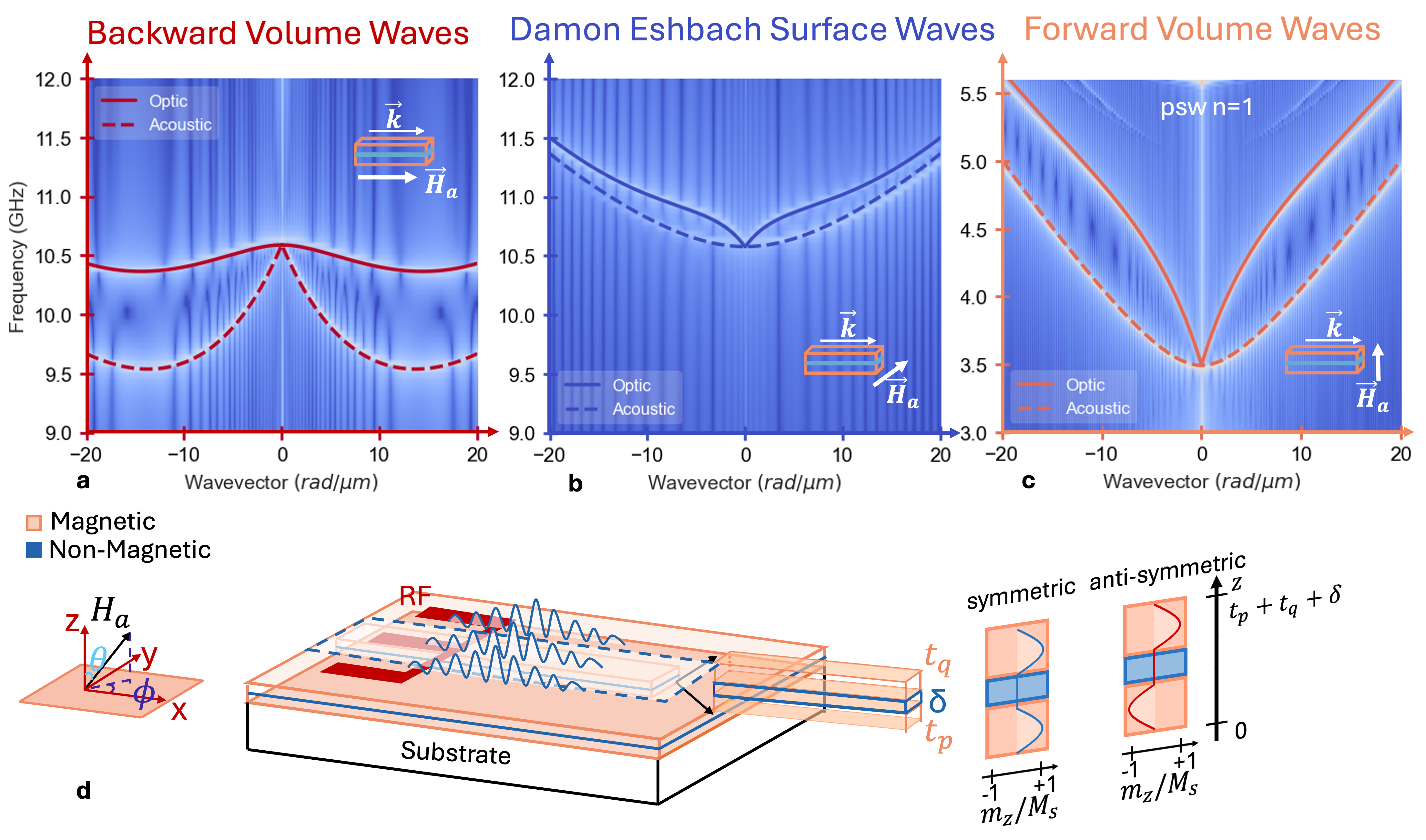}
\caption{
\label{fig:SIM}\textbf{a–c}, Comparison between micromagnetic simulations performed using OOMMF (see Supplementary S3) and the analytical model for the backward-volume, Damon–Eshbach, and forward-volume configurations, respectively, in the absence of magnetic anisotropy, for YIG 100 nm/NM 4 nm/ YIG 100 nm thin film hetero-structure with a  $200x200\ \mu$m  simulation size, with $H_a=300$ mT. \textbf{d} Schematic of the system and reference frames, $\theta$ being the polar angle and $\phi$ the azimuthal one. Two magnetic layers of thickness $t_p$ and $t_q$, separated by a non-magnetic spacer of thickness $\delta$, support symmetric and antisymmetric oscillations under radio-frequency excitation.
}
\end{figure*}

\subsection{\label{subsec:NO_AN}Homogeneous saturation magnetization in the absence of anisotropy}

To extract the essential physical features of dipolarly coupled systems, we focus the analysis to two identical magnetic layers separated by a non-magnetic spacer, neglecting magnetic anisotropies (Fig. \ref{fig:SIM}). For these cases, it is possible to derive a closed and compact analytical formulation that highlights the role of dynamic dipolar coupling.

The diagonal blocks of the interaction tensor satisfy $\hat{\boldsymbol{\Omega}}^{(11)}=\hat{\boldsymbol{\Omega}}^{(22)}$, while the mutual interaction blocks obey $\hat{\boldsymbol{\Omega}}^{(12)}\neq\hat{\boldsymbol{\Omega}}^{(21)}$. This asymmetry originates directly from the opposite signs of the off-diagonal demagnetizing tensor components $\hat{\boldsymbol{N}}_{k,xz}^{(pq)}$ and $\hat{\boldsymbol{N}}_{k,xz}^{(qp)}$. The physical origin of this behavior lies in the layered geometry: the presence of an upper magnetic layer introduces a directional dependence in the dipolar interaction that is absent in laterally coupled systems \cite{doi:10.1126/sciadv.1701517}, where the dipolar tensor remains symmetric, i.e., $\hat{\boldsymbol{\Omega}}^{(pq)}=\hat{\boldsymbol{\Omega}}^{(qp)}$.

In vertically coupled hetero-structures, the relative displacement along the out-of-plane direction introduces a nontrivial phase factor in the mutual dipolar interaction. In particular, this results in a $\pi$-phase shift between the anti-diagonal components of the interaction tensor $\hat{{\Omega}}_c$. This distinction can be formally understood by examining the construction of the diagonal and off-diagonal tensor blocks. The diagonal terms $\hat{\boldsymbol{\Omega}}^{(pp)}$ arise from integration over the thickness of each individual waveguide and therefore describe intralayer interactions, remaining formally analogous to the lateral case. In contrast, the mutual terms $\hat{\boldsymbol{\Omega}}^{(pq)}$ ($p\neq q$) involve integration across the full hetero-structure thickness and encode interlayer dipolar coupling. This structural asymmetry is explicitly captured by the auxiliary functions $\Delta_{pq}$ and $\tilde{\Delta}_{pq}$, which govern self- and mutual-layer contributions respectively, and the sign function $\mathrm{sign}(z_{p}-z_{q})$ present in $xz$ and $yz$ terms of $\hat{\mathbf{N}}_{k}^{(pq)}$.

Despite the lack of symmetry between the mutual interaction tensors, the mathematical properties of the integrals defining the demagnetizing tensor allow a partial simplification of the problem. In particular, it leads to:
\begin{itemize}
    \vspace{0.1pt}
    \item$F_{k_x,xy}^{(12)}=-F_{k_x,xy}^{(21)}$, with $F_{k_x,xy}^{(12)},F_{k_x,xy}^{(21)}\in\mathbb{R}$,\vspace{0.1pt}
    \item$F_{k_x,xz}^{(12)}=-F_{k_x,xz}^{(21)}$, with $F_{k_x,xz}^{(12)},F_{k_x,xz}^{(21)}\in\mathbb{I}\mathrm{m}$,\vspace{0.1pt}
    \item$F_{k_x,ii}^{(12)}=F_{k_x,ii}^{(21)}$, with $F_{k_x,ii}^{(12)},F_{k_x,ii}^{(21)}\in\mathbb{R}$,
    
\end{itemize}
where $i=x,y,z$.
This approach enables a unified treatment of spin-wave dispersion according to the relative angle between the magnetization and the applied external magnetic field, leading to the general mathematical form:
\begin{equation}
    \omega_{s,as}=\sqrt{\Omega_0\pm\omega_M\sqrt{\Omega'}}.
\end{equation}
In each case, the unit vector $\vec{\mu}$ defines the equilibrium magnetization direction, and its substitution into the skew-symmetric matrix $\hat{\bm{\mu}}_C$ allows the derivation of the respective analytical expressions. Here, we explicitly state the analytical forms for the three main dipolar configurations;
\begin{widetext}
\begin{itemize}

\item \textit{Backward-volume (BV) waves} ($\vec{k}\parallel\vec{\mu}$, coplanar):
\begin{equation}
\begin{aligned}
    \label{eq:analyticalSOL_BV}
\Omega_{0}&=\Omega_{yy}\Omega_{zz}+\omega_{M}^{2}F_{yy}^{(12)}F_{zz}^{(12)},\\
    \Omega'&=\left(\Omega_{yy}F_{zz}^{(12)}+\Omega_{zz}F_{yy}^{(12)}\right)^{2}.
\end{aligned}
\end{equation}

\item \textit{Damon--Eshbach (DE) surface waves } ($\vec{k}\perp\vec{\mu}$, coplanar):
\begin{equation}
 \label{eq:analyticalSOL_DE}
\begin{aligned}
    \Omega_{0}&=\Omega_{xx}\Omega_{zz}+\omega_{M}^{2}F_{xx}^{(12)}F_{zz}^{(12)}+\omega_{M}^{2}\left(F_{xz}^{(12)}\right)^{2},\\
    \Omega'&=\left(\Omega_{zz}F_{xx}^{(12)}+\Omega_{xx}F_{zz}^{(12)}\right)^{2}
    +4\omega_{M}^{2}F_{xx}^{(12)}F_{zz}^{(12)}\left(F_{xz}^{(12)}\right)^{2}.
\end{aligned}
\end{equation}

\item \textit{Forward-volume (FV) waves} (non coplanar $\vec{k}$ and $\vec{\mu}$):
\begin{equation}
\label{eq:analyticalSOL_FV}
\begin{aligned}
    \Omega_{0}&=\Omega_{xx}\Omega_{yy}+\omega_{M}^{2}F_{xx}^{(12)}F_{yy}^{(12)}
    +\omega_{M}^{2}\left(F_{xy}^{(12)}\right)^{2}-\omega_{M}^{2}F_{xy}^{2},\\
    \Omega'&=\left(\Omega_{yy}F_{xx}^{(12)}+\Omega_{xx}F_{yy}^{(12)}\right)^{2}
    +4\omega_{M}^{2}F_{xx}^{(12)}F_{yy}^{(12)}\left(F_{xy}^{(12)}\right)^{2}
    -4\omega_{M}^{2}F_{xy}^{2}\left(F_{xy}^{(12)}\right)^{2}.
\end{aligned}
\end{equation}
\end{itemize}

\end{widetext}

\begin{figure*}

\includegraphics[scale=0.5]{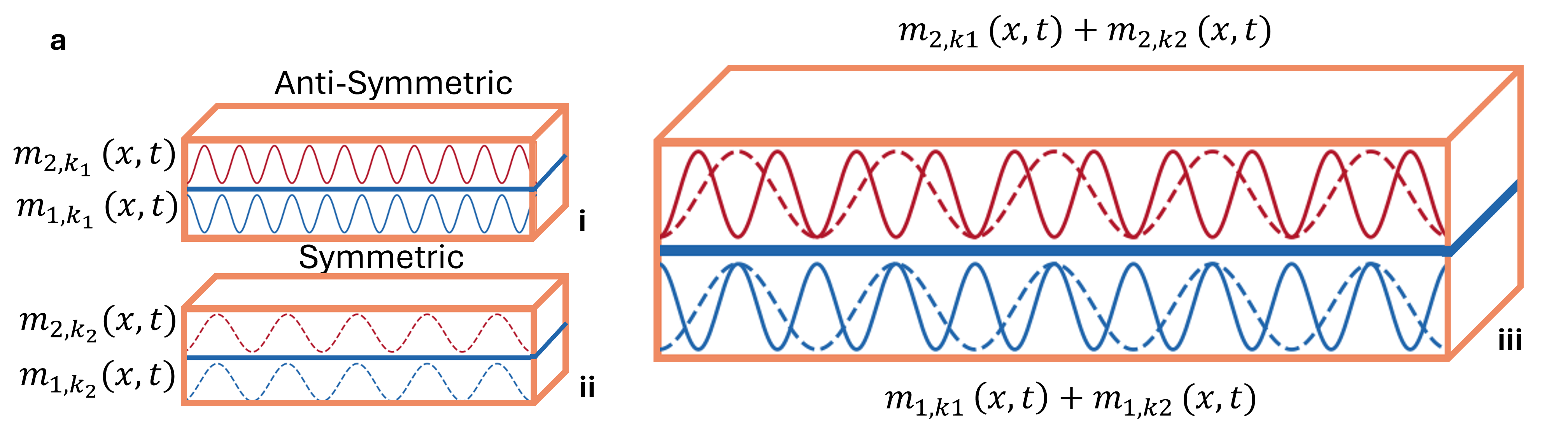}
\includegraphics[scale=0.5]{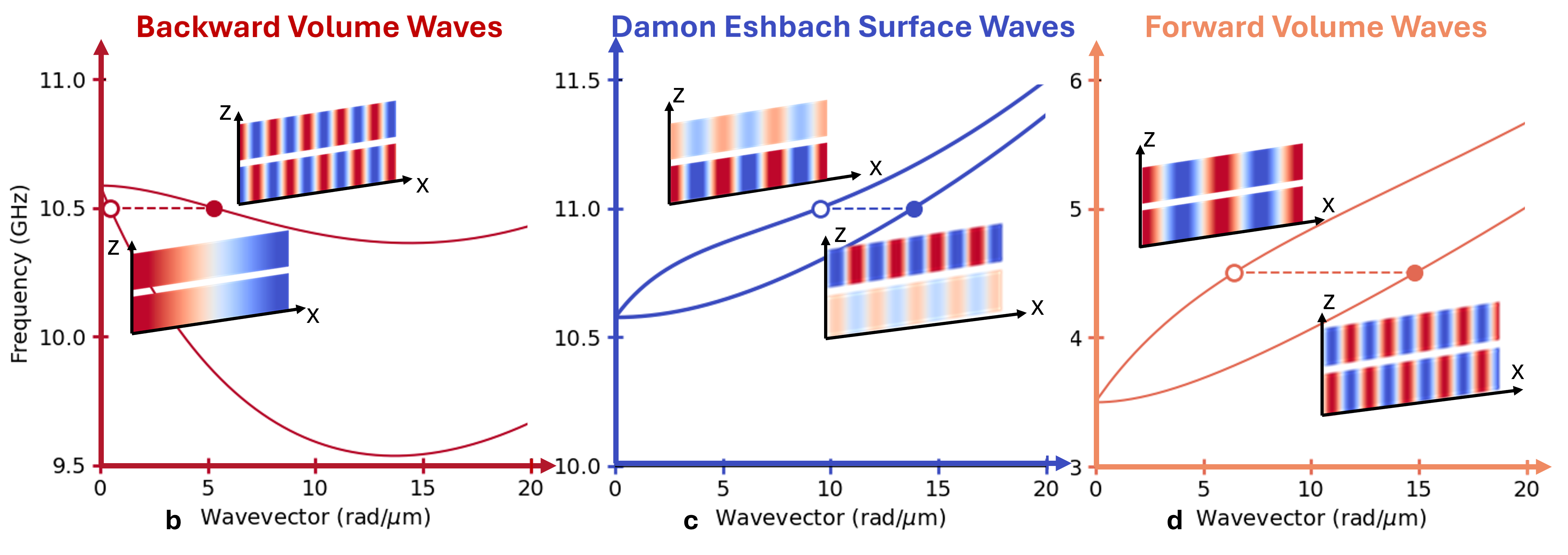}
\includegraphics[scale=0.5]{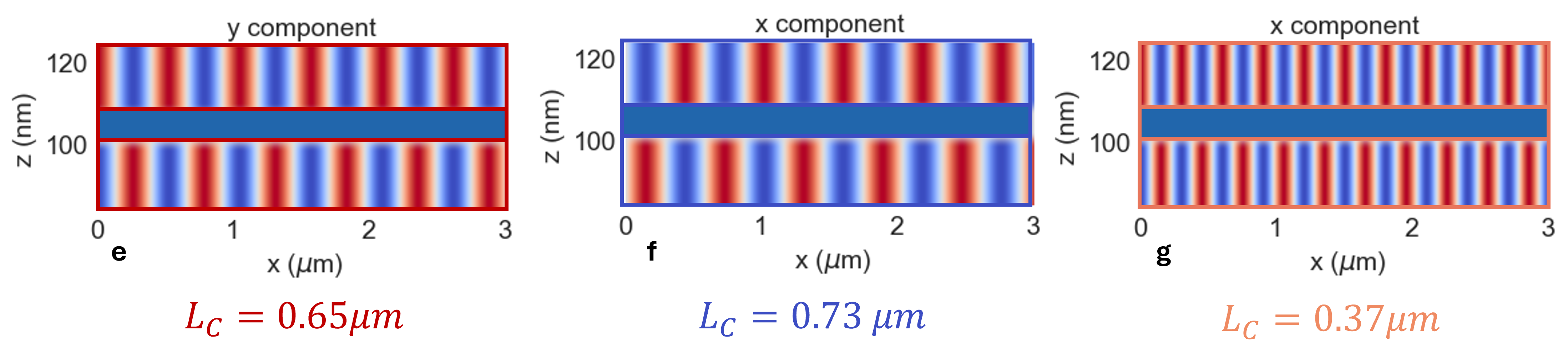}
\caption{\label{fig:EIG}
\textbf{a} Qualitative sketch of the symmetric and anti-symmetric spin waves profiles (i,ii) and their collective behavior due to interference at fixed frequency (iii).
\textbf{b–d}, Dispersion relations in the positive-wavevector region, highlighting the symmetric and antisymmetric character of the oscillating magnetization at the resonant frequency for each spin-wave configuration. Insets show the associated spin-wave mode normalized profile amplitude ($m$ unit vector) at fixed wavector in both acoustic and optical branch.
\textbf{e–g}, Normalized oscillating spatial intensity (cos$(\Delta k_x+\Delta \phi_k)$) of the dynamic magnetization resulting from the coexistence of the two wavevectors selected at a fixed excitation frequency (as indicated in b-d dashed lines), shown for the BV , DE, and FV configurations , respectively, together with the associated coupling length $ L_C$.
}
\end{figure*}

Upon numerical evaluation of the integrals defining the tensor elements $\left(\hat{\boldsymbol{F}}_{k_x}^{(pq)}\right)_{ij}$, the corresponding dispersion relations can be obtained. The resulting spectra, shown in Fig.~\ref{fig:SIM} a-c, display good agreement with OOMMF micromagnetic simulation \cite{beg2022}, and TetraX \cite{TetraX,korberFiniteelementDynamicmatrixApproach2021a} packages (see Supplementary S3). From the eigenvalue solution, the associated eigenvectors can be obtained, allowing the reconstruction of the spatial profiles of the dynamic magnetization, as illustrated in Fig.~\ref{fig:EIG} b-d. When the system is driven at a fixed frequency, the excitation intersects both the symmetric and antisymmetric branches at the wavevectors $k_s$ and $k_{as}$. The resulting superposition of these modes gives rise to the spatial modulation of the spin-wave intensity across the hetero-structure .

For a given magnetic layer $p$, and component $i$, the specific dynamic magnetization component can be written explicitly as:

\begin{equation}
\begin{aligned}
\tilde{m}(x,t) =\;&
2 \lvert m_{k_s} \rvert
\cos\!\left(k_s x + \phi_{k_s} - \omega t\right) \\
&+
2 \lvert m_{k_{as}} \rvert
\cos\!\left(k_{as} x + \phi_{k_{as}} - \omega t\right)
\end{aligned}
\end{equation}
where $\lvert m_{k_s} \rvert$, $\lvert m_{k_{as}} \rvert$ and 
$\phi_{k_s}$, $\phi_{k_{as}}$ are the moduli and phases of the respective
eigenvectors. This leads to an associated time-averaged intensity
$I(x) \propto \langle \tilde{m}(x,t) \rangle_t$:

\begin{equation}
I(x) =
\lvert m_{k_s} \rvert^2
+
\lvert m_{k_{as}} \rvert^2
+
\lvert m_{k_s} \rvert
\lvert m_{k_{as}} \rvert
\cos\!\left(\Delta k_x\, x + \Delta \phi_{k}\right),
\label{eq:15}
\end{equation}

showing that the resulting interference pattern is sinusoidally distributed along the propagation direction, periodically concentrating the energy into one or the other layer, as depicted in \ref{fig:EIG} e-g.
For each wavevector $k$, the dispersion relation exhibits both  optical and acoustic branches, i.e. available modes at higher and lower frequencies, respectively. These correspond to collective symmetric and antisymmetric excitations of the hetero-structure, as can be appreciated from the reported spatial profiles in Fig.~\ref{fig:EIG}.

Even within this simplified setting, we emphasize that the analytical model is built upon the linearization of the Landau--Lifshitz equation, where the two magnetic layers are initially treated as independent subsystems, as reflected in the structure of the tensor $\hat{\boldsymbol{\Omega}}_{k_x}^{(pq)}$ on the right-hand side of Eq.~\eqref{eq:TensorForm}.

The dynamic coupling between the layers emerges only through the careful evaluation of the coupled terms. In fact recasting the right-hand side of eq.\ref{eq:LL_linear} with the auxiliary skew matrix properly captures the complex dipolar landscape of the structure. It describes both the behavior of the single layer as an isolated system, and the energetic terms from being embedded in the hetero-structure, thus preventing a simple linear Bloch-type picture of the form $\Omega_{0}\pm\Omega'$. This highlights the intrinsically nontrivial nature of the modes in dipolarly coupled systems, where the eigenfrequencies result from the full tensorial structure of the mutual interaction rather than from a perturbative splitting of uncoupled modes.
Figure~\ref{fig:EIG} b-d schematically illustrates the spatial profiles associated with the symmetric and antisymmetric modes. The two precession states exhibit the same resonant frequency with distinct spatial modulation. In principle, this can be exploited to induce controlled interference effects within the hetero-structure through appropriate design of the radio-frequency transducers \cite{vlaminck2010spin,qin2018propagating}.

\section{\label{sec:Perturbation}Simplified perturbative approach}

As discussed above, a full analytical determination of the eigenvalues and eigenvectors of the system requires solving the complete coupled spin-wave eigenproblem,
\begin{equation}
    \hat{\boldsymbol{\mu}}_{C}\hat{\boldsymbol{\Omega}}_{C}\,\mathbf{m}
    =
    -\mathrm{i}\omega\,\mathbf{m},
    \label{eq:eig_pert}
\end{equation}
where the intrinsic coupling between the layers is encoded in the tensor $\hat{\boldsymbol{\Omega}}$. While this formulation captures the physics exactly, its complexity motivates the introduction of a simplified perturbative approach, which nevertheless provides insight into the nature and energy of the collective modes.

In the  perturbative regime the eigenvalues of the uncoupled layers are obtained from $H^{0}$, while the effect of interlayer dipolar coupling is treated as a perturbation through $H'$. 

So  Eq.~\eqref{eq:eig_pert} becomes:
\begin{equation}
\left( H^{0} + \lambda H' \right)
\left( \psi + \lambda \psi' \right)
=
-\mathrm{i}
\left( \omega_{0} + \lambda \omega' \right)
\left( \psi + \lambda \psi' \right),
\label{eq:1stord_exp}
\end{equation}
and noting that in the BV, DE, and FV configurations, the full operator $\hat{\boldsymbol{\mu}}_{C}\hat{\boldsymbol{\Omega}}_{C}$ can always be reduced to a four-dimensional subspace, the eigenproblem can be written as
\begin{equation}
\left[
\begin{pmatrix}
H_{pp}^{0} & 0 \\
0 & H_{qq}^{0}
\end{pmatrix}
+
\begin{pmatrix}
0 & H'_{pq} \\
H'_{qp} & 0
\end{pmatrix}
\right]
\begin{pmatrix}
c_{1} \\
c_{2}
\end{pmatrix}
\mathbf{m}
=
-\mathrm{i}\omega
\begin{pmatrix}
c_{1} \\
c_{2}
\end{pmatrix}
\mathbf{m}.
\end{equation}

Since the problem is non-Hermitian, left and right eigenvectors must be introduced, satisfying the biorthonormality condition $\langle \phi_i^{L} | \phi_j^{R} \rangle = \delta_{ij}$ \cite{li2025_PERT}.
\\
The perturbative eigenproblem then takes the generalized form
\begin{equation}
W
\begin{pmatrix}
\alpha \\
\beta
\end{pmatrix}
=
\begin{pmatrix}
\langle \phi_{1}^{L} | H' | \phi_{1}^{R} \rangle
&
\langle \phi_{1}^{L} | H' | \phi_{2}^{R} \rangle
\\
\langle \phi_{2}^{L} | H' | \phi_{1}^{R} \rangle
&
\langle \phi_{2}^{L} | H' | \phi_{2}^{R} \rangle
\end{pmatrix}
\begin{pmatrix}
\alpha \\
\beta
\end{pmatrix}
=
-\mathrm{i}\omega'
\begin{pmatrix}
\alpha \\
\beta
\end{pmatrix},
\end{equation}
with $W$ the perturbative matrix of the degenerate states.
This system yields two degenerate unperturbed solutions at frequency $\omega_0$, associated with left and right eigenvectors $\phi^{L,R}_{1,2}$. A compact expression, valid for all three configurations in an infinite film, is obtained by choosing $(i,j)=(y,z)$ in the BV case, $(x,z)$ in the DE case, and $(x,y)$ in the FV case:
\begin{equation}
\begin{aligned}
\phi_{1}^{L}
&=
\frac{1}{\sqrt{2}}
\left(
\pm \mathrm{i} \frac{\Omega_{ii}}{\omega_{0}},
\; 1,
\; 0,
\; 0
\right),
\qquad
\phi_{1}^{R}
=
\frac{1}{\sqrt{2}}
\begin{pmatrix}
\mp \mathrm{i} \dfrac{\Omega_{jj}}{\omega_{0}} \\
1 \\
0 \\
0
\end{pmatrix},
\\[0.6em]
\phi_{2}^{L}
&=
\frac{1}{\sqrt{2}}
\left(
0,
\; 0,
\; \pm \mathrm{i} \frac{\Omega_{ii}}{\omega_{0}},
\; 1
\right),
\qquad
\phi_{2}^{R}
=
\frac{1}{\sqrt{2}}
\begin{pmatrix}
0 \\
0 \\
\mp \mathrm{i} \dfrac{\Omega_{jj}}{\omega_{0}} \\
1
\end{pmatrix}.
\end{aligned}
\end{equation}

Projecting the perturbation onto the degenerate subspace yields
\begin{equation}
W
=
\begin{pmatrix}
0
&
-\dfrac{\mathrm{i}\omega_{M}}{2}\,\Delta
\\[0.4em]
-\dfrac{\mathrm{i}\omega_{M}}{2}\,\Delta
&
0
\end{pmatrix},
\qquad
\Delta
=
\frac{
\Omega_{ii} F_{jj}
+
\Omega_{jj} F_{ii}
}{\omega_{0}},
\end{equation}
which allows the calculation of the frequency correction $\omega'$ and the associated eigenvectors $\psi^{L,R}_{\pm}=\alpha_{\pm}\phi^{L,R}_{1}+\beta_{\pm}\phi^{L,R}_{2}$:
\begin{equation}
\begin{aligned}
\omega'
&=
\pm \frac{\omega_{M}}{2}\lvert \Delta \rvert,
\\[0.6em]
\psi_{+}^{R}
&=
\begin{pmatrix}
\dfrac{\Delta}{\lvert \Delta \rvert} \\
1
\end{pmatrix},
\qquad
\psi_{-}^{R}
=
\begin{pmatrix}
-\dfrac{\Delta}{\lvert \Delta \rvert} \\
\,\,\,\,\, 1
\end{pmatrix}.
\end{aligned}
\end{equation}

\begin{figure}
\includegraphics[width=0.5\linewidth]{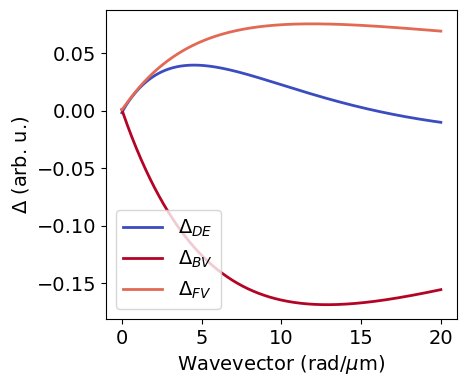}
\caption{\label{fig:DELTA-EIGENVECTOR}Frequency correction, $\Delta$, calculated  from the simplified perturbative model for different spin-waves configurations in the case of $w=100\ \mu\mathrm{m}$. Notably, the BV configuration exhibits an opposite trend compared to DE and FV, accounting for the distinct association of symmetric and antisymmetric modes with the acoustic and optical branches.}
\end{figure}

These solutions represent the collective symmetric and antisymmetric oscillations of the two magnetic layers. Within this framework, it is possible to identify how the mode symmetry is associated with the acoustic or optical branches at a given wavevector. Inspection of Fig. \ref{fig:DELTA-EIGENVECTOR} reveals that the sign of $\Delta$ exhibits opposite behavior in the backward-volume configuration (BV) compared to the Damon-Eshbach (DE) and forward-volume (FV) cases. As a consequence, the symmetric and antisymmetric modes belong to the optical and acoustic branches respectively in the DE and FV configurations, while the opposite occurs in the BV configuration. This implies that, at a fixed excitation frequency, the symmetric mode is associated with the smaller wavevector and therefore with the larger wavelength. Within this simplified perturbative approach, the separation between the bands and its behaviour in laterally confined cases can be directly evaluated and compared with the computed $\Omega\,'$ from Eqs. \eqref{eq:analyticalSOL_BV} , \eqref{eq:analyticalSOL_DE} , \eqref{eq:analyticalSOL_FV}. (See Supplementary S4)

\section{\label{sec:Model_evaluation}Model evaluation}

Having established the analytical framework, we now evaluate the predictive capabilities of the model and illustrate how appropriate choices of geometry and material parameters allow engineering of the spin-wave dispersion relations and the associated spin-wave modes. We begin by considering common representative magnonic materials: yttrium iron garnet (YIG), permalloy (NiFe), and cobalt iron boron (CoFeB), in BV and DE spin-wave configurations \cite{szulc_magnetic_2025,doi:10.1126/sciadv.1701517,MS-qin_nanoscale_2021,zenbaa2025yig,BIL-odintsov2022nonreciprocal,REIGBIL-jin2023interface,REIGBIL-rathore2026magnon}. The corresponding materials parameters and film thicknesses used in the calculations are summarized in Table~\ref{tab:MM param}.

\begin{figure*}

\includegraphics[scale=0.5]{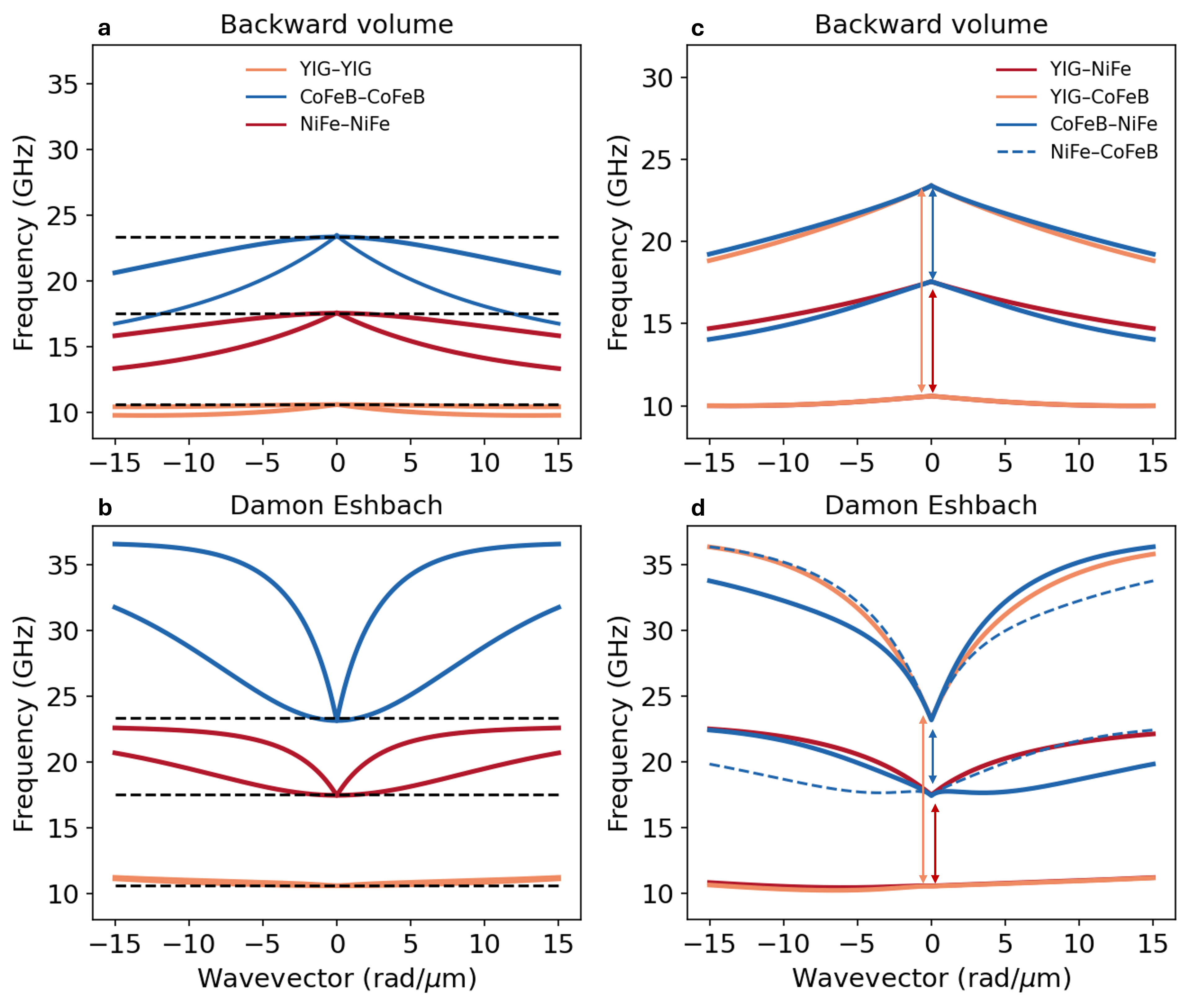}
\caption{\label{fig:Band-mat}
Calculated spin-wave dispersion relations in the backward-volume and Damon–Eshbach configurations for M/NM/M trilayers composed of the same magnetic material : YIG/NM/YIG, CoFeB/NM/CoFeB, and NiFe/NM/NiFe (\textbf{a,b}). All structures have identical magnetic layer thicknesses $t_p=t_q=t=80$ nm and a spacer thickness $\delta=20$ nm. The dashed black lines indicate the coincident FMR frequencies of the two magnetic layers in the homogeneous case.
\textbf{c,d} Spin-wave dispersion relations in the BV and DE configurations, respectively, for  trilayers containing different magnetic materials: YIG/NM/NiFe, YIG/NM/CoFeB, and NiFe/NM/CoFeB, as well as the inverted stacking order CoFeB/NM/NiFe. The layer thicknesses are $t_p=t_q=t=80$ nm and $\delta=20$ nm. These panels highlight the impact of material asymmetry on the separation (vertical arrows) between the spin-wave branches and on the emergence of non-reciprocal behaviour in the DE configuration. 
}

\end{figure*}

In Fig.~\ref{fig:Band-mat}, the resulting dispersion relations for the studied hetero-structures are shown, including symmetric CoFeB, NiFe, and YIG bilayers, as well as their asymmetric combinations.

As expected, the ferromagnetic resonance (FMR, i.e. $k_{x}=0$) coincides in the DE and BV configurations. In contrast, making the two layers from different materials leads to the opening of a gap at $k_{x}=0$ with respect to the case of hetero-structures composed of the same magnetic material (vertical arrows in Fig. \ref{fig:Band-mat} (c,d) ). No dipolar coupling occurs for $k\rightarrow 0$, due to vanishing off-diagonal terms in Eq.~\ref{eq:TensorForm}, i.e., retrieving the expected distinct FMR of the two layers. 
Notably, comparison between the NiFe/CoFeB and CoFeB/NiFe stacks demonstrates that reversing the material order provides an additional degree of freedom to control the propagation direction and group velocity of spin waves.

In general, when the two magnetic layers possess different micromagnetic parameters, a finite gap opens between the acoustic and optical branches. Equivalently, this behavior can be interpreted as the lifting of a degeneracy at $k=0$, as will be discussed in more detail in section A.
The gap opening between the optical and acoustic branches can be directly traced back to the micromagnetic contributions entering the definition of the internal fields, which modify the diagonal terms of the interaction tensor. This indicates that material doping, interface engineering, or growth-induced variations in magnetic parameters can be used to tune the dispersion relations and mode symmetry according to the desired functionality of the hetero-structure.

In particular, in the DE configuration for non-equivalent layers, a pronounced nonreciprocal behavior emerges\cite{DP_Wintz2016-uy,DP_https://doi.org/10.1002/adma.201906439,DP_Girardi2024-el}. In contrast to the BV case, in the DE configuration the $F_{xz}$ and $F_{zx}$ terms enter explicitly into the eigenvalue formulation, thereby introducing a nonreciprocal correction. As will be further clarified in section \ref{subsec:Iron_garnets}, this behavior represents a general property of this micromagnetic configuration, namely that any micromagnetic asymmetry between the two layers is translated into a nonreciprocal spin-wave response.

\begin{table}
\begin{ruledtabular}
\begin{tabular}{ccccccc}
Material&$M_s\,\mathrm{[kA/m]}$&$A\,\mathrm{[pJ/m]}$& $w\,\mu\mathrm{m}$&$t_1=t_2\,\mathrm{[nm]}$&$\delta\,\mathrm{[nm]}$&$H_a\,\mathrm{[mT]}$ \\
\hline

YIG  &$140$&$4$  &200& $100$  & $ 4$ & $300$  \\
NiFe &$800$&$10$ &200& $80$   & $20$ & $300$ \\
CoFeB&$1600$&$15$&200& $80$   & $20$ & $300$\end{tabular}
\end{ruledtabular}
\caption{Micromagnetic and geometric parameters used in the model evaluation.}
\label{tab:MM param}
\end{table}

\begin{figure*}
    \centering
\includegraphics[scale=0.5]{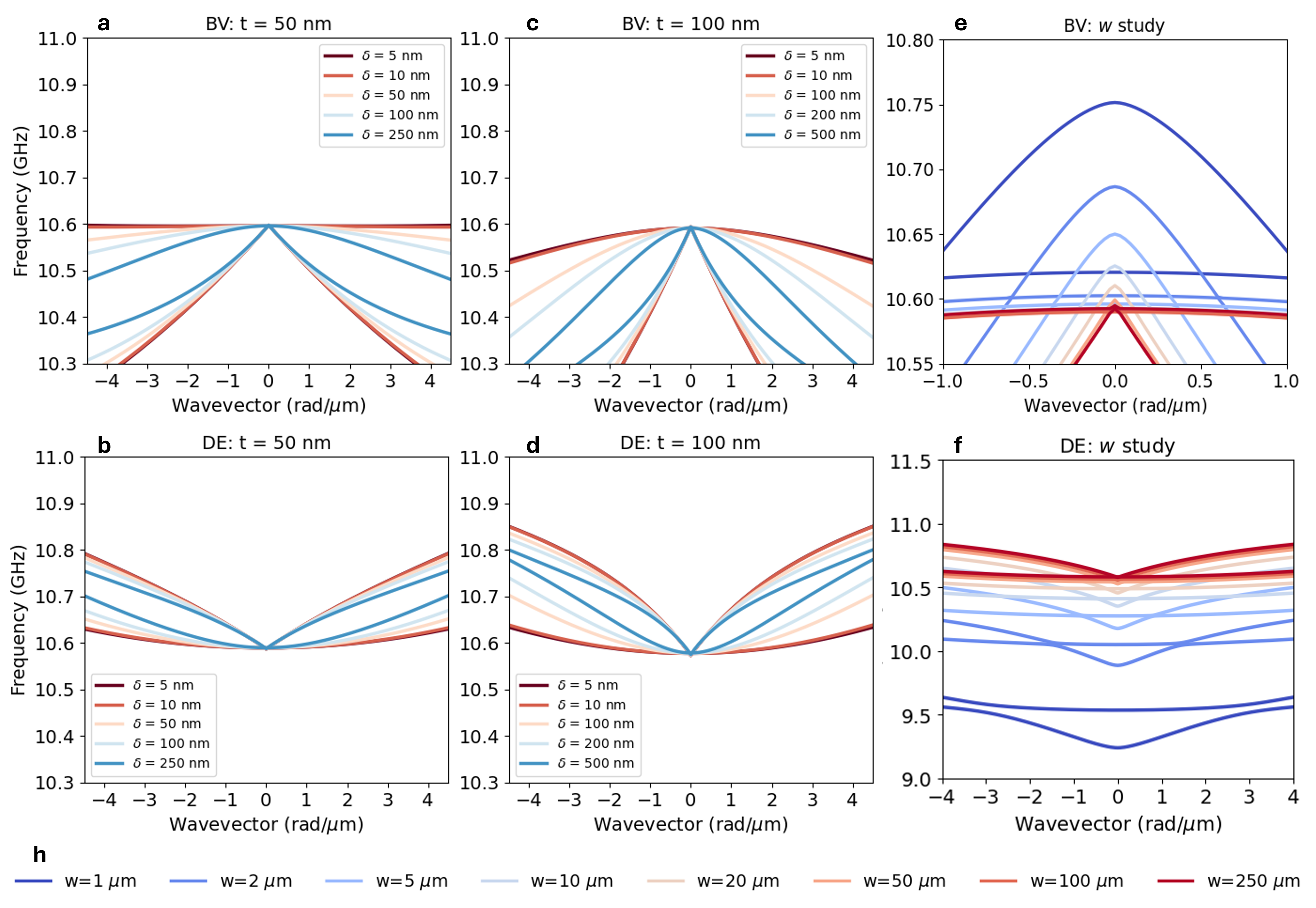}
    
    \caption{\label{fig:d_w}
\textbf{a,b} Calculated spin-wave dispersion relations for two identical infinite YIG layers with thickness $t = 50$ nm (parameters in Table \ref{tab:MM param}), separated by a non-magnetic spacer of increasing thickness $\delta$.
\textbf{c,d} Calculated spin-wave dispersion relations for two identical infinite YIG layers with thickness $t = 100$ nm (parameters in Table \ref{tab:MM param}), separated by a non-magnetic spacer of increasing thickness $\delta$. Increasing the spacer thickness progressively reduces the separation between the acoustic and optic branches, approaching the single-waveguide limit, consistent with the decay of dynamic dipolar coupling with interlayer distance.
\textbf{e,f} Calculated spin-wave dispersion relations of a YIG (100~nm)/NM (4~nm)/YIG (100~nm) waveguide, with parameters reported in Table \ref{tab:MM param}, as a function of the waveguide width (\textbf{h}).}
\end{figure*}


\subsection{\label{subsec:Iron_garnets}Dipolarly coupled iron garnets}

We now focus our analysis on iron-garnet-based hetero-structures, as these materials provide low damping as well as tunability of magnetization and anisotropy, and can be grown as high quality epitaxial hetero-structures. \cite{ANI-jcs6070203, ANIS-bauer2020dysprosium, ANIS-fakhrul2023substrate,ANIS-song2023engineering, ANIS-song2024temperature,ANIS-rosenberg2021magnetic,NON-RECfakhrul2019magneto}. 
For instance, in pulsed-laser-deposited (PLD) films, not only can the cation-site occupancy be tailored through chemical composition, but growth parameters such as substrate choice, deposition atmosphere, and temperature can be exploited to engineer strain and anisotropy \cite{AlisonNature,SITE-rosenberg2023revealing}. Perpendicular magnetic anisotropy can be induced via magnetoelastic or magnetotaxial effects and tuned through lattice or thermal mismatch strain between the film and the substrate\cite{SUB-yoshimoto2018static,SUB-fakhrul2023influence,HET-YIGAuYIG-wu2018magnon}.

Moreover, an appropriate choice of the non-magnetic spacer could allow the suppression of interfacial exchange coupling, while preserving the epitaxy, and then ensuring that the interaction between the magnetic layers is purely dipolar. From a modeling perspective, iron garnets therefore represent a nearly ideal realization of the assumptions underlying the analytical framework developed in this work \cite{HET-barman20212021,HET-https://doi.org/10.1155/2012/168313,HET-li2020magnon,HET-TmMgoTm-https://doi.org/10.1002/adfm.202526781,HET-YIGAuYIG-wu2018magnon,HET-YIGNiO-YIG-guo2018magnon,HET-YIGPtYIG-truong2024optimization}. Their intrinsically low damping enables a direct and reliable comparison between analytical predictions and micromagnetic simulations, while the absence of strong interfacial exchange coupling isolates the dipolar interaction as the dominant coupling mechanism. As a result, garnet-based hetero-structures provide a clean and experimentally relevant testbed to assess the predictive power of the model and to explore how controlled layer asymmetry influences collective spin-wave dispersion and mode behavior\cite{ANI-jcs6070203,ANIS-bauer2020dysprosium,ANIS-rosenberg2021magnetic,ANIS-song2023engineering}.

We start by considering the  simplified case of two identical YIG layers separated by a non-magnetic spacer. In Fig.~\ref{fig:d_w}(a-d), the impact of the YIG layer thickness and spacer thickness on the spin-wave dispersion is presented.
\begin{figure*}

\includegraphics[scale=0.5]{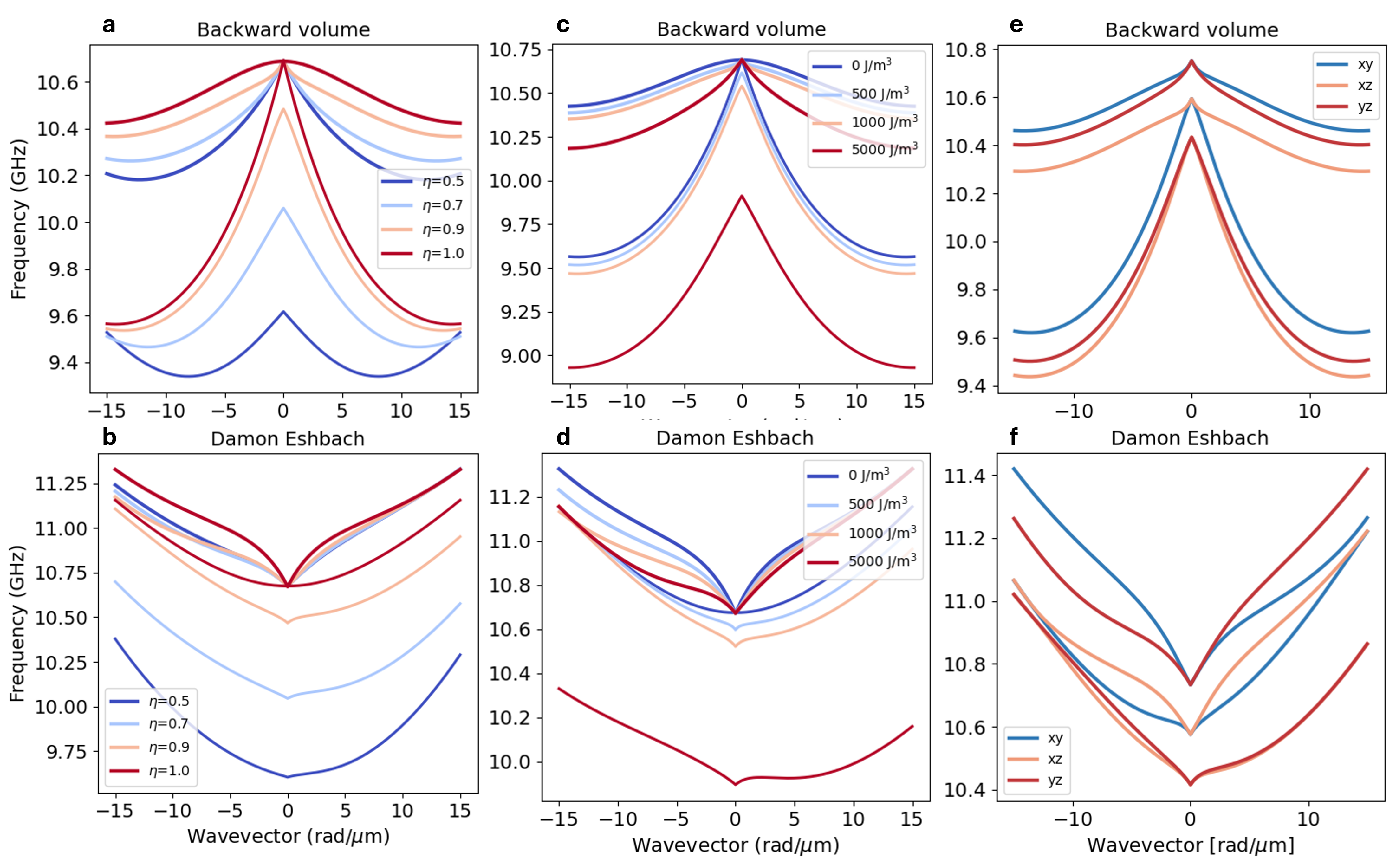}
\caption{\label{fig:Ms,K}
\textbf{a,b} Spin-wave dispersion relations in the backward volume  and Damon-Eshbach configurations for two YIG layers with different saturation magnetizations, where $M_{s,q}=\eta M_{s,p}$ and $\eta\leq1$, in the absence of magnetic anisotropy. As the mismatch between the saturation magnetizations of the two layers increases, the separation between the acoustic and optic branches increases, with the acoustic branch shifting to lower frequencies.\textbf{c,d} Spin-wave dispersion relations in the BV and DE configurations for two YIG layers with identical saturation magnetization, where the top layer hosts a perpendicular magnetic anisotropy modeled as a uniaxial anisotropy along $\mathbf{u}=(0,0,1)$. Results are shown for different values of the anisotropy constant $K$. The introduction of anisotropy produces an asymmetry between the layers, leading to an increased separation between the branches, reproducing the same qualitative trend observed in panels \textbf{a,b}. In the DE configuration, this anisotropy-induced asymmetry modifies the dispersion and the spatial distribution of the dynamic magnetization.\textbf{e,f} Spin-wave dispersion relations in the BV and DE configurations for two YIG layers with identical saturation magnetization and an anisotropy constant $K = 1000$ J\,m$^{-3}$. The panels illustrate the effect of crossed anisotropy directions between the two layers. Specifically, configurations with the bottom layer anisotropy axis $\mathbf{u}=(1,0,0)$ and the top layer axis $\mathbf{u}=(0,1,0)$ (xy crossed anisotropy), as well as mixed in-plane/out-of-plane (x or yz) configurations with $\mathbf{u}=(1,0,0)$ or $(0,1,0)$ in the bottom layer and $\mathbf{u}=(0,0,1)$ in the top layer, are considered.}
\end{figure*}
At a fixed spacer thickness $\delta$, increasing the magnetic layer thickness leads to an increase in the group velocity. Conversely, for increasing spacer thickness, the acoustic-optical gap progressively decreases, approaching the limiting case of a single, isolated layer.
In Fig.~\ref{fig:d_w}(e,f), an equivalent analysis is performed by varying the width of the YIG/NM/YIG stack, spanning from the infinite-film limit, where the degeneracy occurs at $k=0$, to laterally confined waveguides, where the degeneracy is shifted to finite wavevectors. Before discussing the presence of a degeneracy point, where the acoustic and optical branch cross, and its geometrical implications in detail, it is important to note that higher-order transverse modes associated with the finite width of the waveguide are neglected here. Despite this simplification, the model still allows the extraction of relevant physical trends.
As can be seen, in both the Damon-Eshbach and backward-volume configurations, decreasing the waveguide width shifts the degeneracy point toward larger wavevectors. In addition, the dispersion branches are shifted in opposite directions depending on the configuration: in the BV geometry, the bands are shifted upward, as shape anisotropy favors magnetization alignment parallel to the propagation direction, while in the DE geometry the dispersion is shifted downward, since shape anisotropy makes the $y$ direction a hard axis. As the waveguide width is reduced, the degenerate point at which the symmetric and antisymmetric wavevectors coincide ($k_s = k_{as}$ at fixed frequency) is progressively pushed to larger wavevectors. For sufficiently narrow waveguides, an additional lifting of this degeneracy is observed when the lateral confinement becomes comparable to the total thickness of the hetero-structure (Fig. \ref{fig:d_w}). 
We next consider the impact of layer asymmetry by introducing different saturation magnetizations $M_s$ for the top and bottom layers at fixed width\cite{MS-qin_nanoscale_2021,REIGBIL-rathore2026magnon}, as well as the effect of an interfacial magnetoelastic term. The latter can be modeled as an additional perpendicular magnetic anisotropy (PMA) in the top layer. The corresponding results are shown in Fig.~\ref{fig:Ms,K}. 

The saturation magnetization $M_s$ has a pronounced impact on both the overall bandwidth of the dispersion relations and the resulting group velocities. In particular, lowering $M_s$ of one layer leads to a systematic shift of the spin-wave branches toward lower frequencies and enhances the separation between the symmetric and antisymmetric modes at low $k$. 

This analysis reveals the range of dispersion relations, gap opening and nonreciprocal behavior that can be obtained by modifying the magnetic properties and geometry of the two layers. \cite{SAF3-gallardo2019reconfigurable}.
In principle, the full anisotropy landscape of each garnet layer can be engineered, for example. Fig.~\ref{fig:Ms,K} presents the effect of varying the anisotropy axis $\vec{u}_p$ of the layers to be orthogonal, altering the nonreciprocal response of the hetero-structure\cite{NON-RECfakhrul2019magneto}. These configurations highlights how anisotropic asymmetries between the layers modifies the frequency separation between the acoustic and optic branches, and can affect the dispersion reciprocity. From a mathematical standpoint, this behavior originates from the introduction of off-diagonal nonreciprocal terms proportional to $K_u$, independently of the micromagnetic configuration, as can be directly inferred from Eq.~\ref{eq:Omega_tensor}. 
\section{\label{sec:Conclusion}Conclusion}
We developed a physical model describing spin-wave dynamics in magnetic/non-magnetic/magnetic hetero-structures and benchmarked its predictions against well established micromagnetic simulation platforms. The model captures the essential features of dipolarly coupled spin-wave modes in vertically stacked systems, providing direct access to their dispersion relations, mode hybridization, and symmetry properties.
We show how the intrinsic tunability of iron garnets, arising from compositional variability, strain-induced effects, and controllable anisotropy landscapes, could be exploited for systematic tailoring of the collective spin-wave spectrum within the proposed theoretical framework. 

The possibility of independently varying the magnetic properties of the two layers, while retaining dipolar coupling as the dominant interaction mechanism, establishes garnet-based hetero-structures as a particularly suitable platform for modeling and design studies of collective magnonic phenomena.
Beyond fundamental interest, the presented framework provides a predictive and physically transparent tool for the design of functional magnonic elements. 

\section*{Data Availability Statement}
\noindent The data that support the findings of this study are available from the corresponding author upon reasonable request.
\section*{Acknowledgements}
The authors acknowledge valuable discussions with R. Bertacco and F. Maspero. L. Menna acknowledges G. Gubbiotti and M. Madami for valuable discussions. O. Wojewoda acknowledges support from a Marie Curie Fellowship. C. A. Ross and M. J. Gross acknowledge support from NSF DMR 2323132. A. Del Giacco acknowledges support from Politecnico di Milano.

\bibliography{aipsamp}
\section*{Supplementary Material}
\subsection*{Linearization of the Landau-Lifschitz Equation in the Reciprocal $k$-space}
To obtain Eq.2 in main text, the starting point is the expression of Landau-Lifshitz (LL) equation in real-space:
\begin{equation}
    \frac{d\vec{M}}{dt}=-\gamma\mu_{0}\left(\vec{M}\times\vec{H}_{eff}\right),
\end{equation}
where the magnetization $\vec{M}$ of the system can be then expressed as:
\begin{equation}
    \vec{M}(\vec{r},t)=M_{S}\vec{\mu}+\vec{m}(\vec{r},t),\quad \vec{m}(\vec{r},t)<<M_{S}\vec{\mu}
\end{equation}
, and $\vec{H}_{eff}$ accounts for the effective field at the equilibrium configuration, also expressed as:
\begin{equation}
    \vec{H}_{eff}=\frac{B}{\mu_{0}}\vec{\mu}+\frac{2A}{\mu_{0}M_{S}^{2}}\boldsymbol{\nabla}^{2}\vec{m}-\hat{\mathbf{N}}(\vec{r}-\vec{r}')\vec{m}(\vec{r}')    
\end{equation}
, where $B=\mu_0(H_a+H_d)$ accounts for the applied external field and the static component of the dipolar field, and the exchange field only depends on $\vec{m}$, since $\vec{\mu}$ has no spatial dependence.

\noindent In this work the first assumption regards the direction of the static effective field (and then the equilibrium magnetization) with respect to the external field $\vec{H}_{a}$.

Plugging these expressions into the LL equation and keeping only linear terms in $\vec{m}^{(p)}$ we get:
\begin{equation}\label{eq: LL_LINEAR}
\begin{aligned}
\frac{d}{dt}\left[M_{S}\vec{\mu}^{(p)}+\vec{m}^{(p)}\right]&=\\&=\gamma\mu_{0}\left[\left(\frac{B^{(p)}}{\mu_{0}}\vec{\mu}^{(p)}+\underbrace{\frac{2A}{\mu_{0}M_{S}^{2}}}_{l_{exc}^{2}}\boldsymbol{\nabla}^{2}\vec{m}^{(p)}-\sum_{q}\hat{\mathbf{N}}(\vec{r}^{(p)}-\vec{r}^{(q)})\vec{m}^{(q)}\right)\times M_{S}\left(\vec{\mu}^{(p)}+\vec{m}^{(p)}\right)\right]=\\&=\gamma\mu_{0}\left(\frac{B^{(p)}}{\mu_{0}}\vec{\mu}^{(p)}\times\vec{m}^{(p)}+l_{exc}^{2}M_{S}\boldsymbol{\nabla}^{2}\vec{m}^{(p)}\times\vec{\mu}^{(p)}+\vec{\mu}^{(p)}\times M_{S}\sum_{q}\hat{\mathbf{N}}(\vec{r}^{(p)}-\vec{r}^{(q)})\vec{m}^{(q)}\right)= \\&= \vec{\mu}^{(p)}\times \sum_{q}\left[\gamma\mu_{0}\left(\frac{B^{(p)}}{\mu_{0}}-M_{S}l_{exc}^{2}\boldsymbol{\nabla}^{2}\vec{m}^{(p)}\right)\delta_{pq}\hat{\mathbf{I}}+\gamma\mu_{0}M_{S}\hat{\mathbf{N}}(\vec{r}^{(p)}-\vec{r}^{(q)})\right]\vec{m}^{(q)}
\end{aligned}
\end{equation}
Introducing $\omega_{M}=\gamma\mu_{0}M_{S}$ and recalling that $\frac{d}{dt}\vec{\mu}^{(p)}=0$:
\begin{equation}
    \frac{d}{dt}\vec{m}^{(p)}=\vec{\mu}^{(p)}\times\sum_{q}\underbrace{\left[\left(\gamma B^{(p)}-\omega_{M}l_{exc}^{2}\boldsymbol{\nabla}^{2}\vec{m}^{(p)}\right)\delta_{pq}\hat{\mathbf{I}}+\omega_{M}\hat{\mathbf{N}}(\vec{r}^{(p)}-\vec{r}^{(q)})\right]}_{\hat{\boldsymbol{\Omega}}_{pq}}\vec{m}^{(q)}
\end{equation}
Taking the Fourier transform of Eq. \ref{eq: LL_LINEAR}, the expression of Eq. 2 in main text is retrieved:
\begin{equation}
    -i\omega\vec{m}^{(p)}(\vec{k})=\vec{\mu}^{(p)}\times \sum_{q}\hat{\boldsymbol{\Omega}}_{k}^{(pq)}\vec{m}^{(q)}(\vec{k}),
\end{equation}
and the static internal field takes the form also reported in main text (Eq. 4), since the dipolar energy contribution in real space corresponds to:
\begin{equation}
    -\mu_{0}M_{S}\sum_{q}\hat{\mathbf{N}}(\vec{r}^{(p)}-\vec{r}^{(q)})\vec{\mu}^{(q)},
\end{equation}
where $\vec{\mu}$, i.e. the equilibrium direction of magnetization, can be also written as the $\vec{M}(k=0)$ spectral component.
\subsection*{Magnetometric Tensor Derivation: Main Steps}
From :
\begin{equation}
    N_{\alpha\beta}(\vec{r}_{p}-\vec{r}_{q})=\frac{1}{\tilde{w}t_{p}}\int_{-\infty}^{+\infty}\sigma_{p}(\vec{k})\sigma^{*}_{q}(\vec{k})\frac{\partial^{2}}{\partial\alpha\partial\beta}\frac{\mathrm{e}^{i\vec{k}\cdot\left(\vec{r}_{p}-\vec{r}_{q}\right)}}{\kappa^{2}}\frac{d^{3}\vec{\kappa}}{(2\pi)^{3}}\quad\quad \alpha,\beta=x,y,z
\end{equation}
where $\sigma(\vec{k})$ corresponds to the Fourier transform of the magnetization profile, and $\tilde{w}$ is a normalization factor.
Following the same derivation as in \cite{VERBA_NANODOTS}, it is convenient to define $\kappa^2=k^{2}+\kappa_{z}^{2}$, $k^{2}$ being equal to $\kappa_{x}^{2}+\kappa_{y}^{2}$, and $r^{2}=\rho^{2}+z^{2}$, $\rho^{2}$ being equal to $x^{2}+y^{2}$.
This results in:
\begin{equation}
    \hat{\mathbf{N}}(\vec{\rho}-\vec{\rho}')=\int dzdz'\hat{\mathbf{N}}(\vec{r}-\vec{r}')=\frac{1}{\tilde{w}t_{p}}\int_{-\infty}^{+\infty}\frac{d^{2}\vec{k}}{(2\pi)^{2}}\int dzdz'\sigma_{p}(\vec{k})\sigma_{q}^{*}(\vec{k})\frac{\partial^{2}}{\partial\alpha\partial\beta}\mathrm{e}^{i\vec{k}\cdot(\vec{\rho}-\vec{\rho}')}\underbrace{\int_{-\infty}^{+\infty}\frac{d\kappa_{z}}{2\pi}\frac{\mathrm{e}^{i\kappa_{z}(z-z')}}{k^{2}+\kappa_{z}^{2}}}_{I}
\end{equation}

To evaluate $I$ we exploit the residual theorem:
\begin{equation}
    I=\frac{1}{2\pi}\cdot2\pi i \, \mathrm{Res}\left\{\left.\frac{\mathrm{e}^{i(u-u')}}{(u+ik)(u-ik)}\right|_{z=ik}\right\}=\frac{1}{2\pi}\cdot\frac{\pi}{k}\mathrm{e}^{-k\left|z-z'\right|}=\frac{1}{2k}\mathrm{e}^{-k\left|z-z'\right|},
\end{equation}
thus resulting in:
\begin{equation}
    \hat{\mathbf{N}}(\vec{\rho}-\vec{\rho}')=\frac{1}{\tilde{w}t_{p}}\int_{-\infty}^{+\infty}\frac{d^{2}\vec{k}}{(2\pi)^{2}}\int dzdz'\sigma_{p}(\vec{k})\sigma_{q}^{*}(\vec{k})\frac{\partial^{2}}{\partial\alpha\partial\beta}\mathrm{e}^{i\vec{k}\cdot\left(\vec{\rho}-\vec{\rho}'\right)}\frac{1}{2k}\mathrm{e}^{-k|z-z'|}
\end{equation}
Expressing now $\alpha\beta$ matrix terms in case of $\alpha,\beta=x,y$, $\alpha z$ and $zz$ terms:
\begin{align}
    \label{eq:N_REAL_SPACE_XX}
    N_{\alpha\beta}(\vec{r}_{p}-\vec{r}_{q})&=\frac{1}{\tilde{w}t_{p}}\int_{-\infty}^{+\infty}\frac{d^{2}\vec{k}}{(2\pi)^{2}}\int dzdz'\sigma_{p}(\vec{k})\sigma_{q}^{*}(\vec{k})\frac{\kappa_{\alpha}\kappa_{\beta}}{k}\mathrm{e}^{i\vec{k}\cdot\left(\vec{\rho}-\vec{\rho}'\right)}\frac{1}{2}\mathrm{e}^{-k|z-z'|} \\
    \label{eq:N_REAL_SPACE_XZ}
    N_{\alpha z}(\vec{r}_{p}-\vec{r}_{q})&=\frac{1}{\tilde{w}t_{p}}\int_{-\infty}^{+\infty}\frac{d^{2}\vec{k}}{(2\pi)^{2}}\int dzdz'\sigma_{p}(\vec{k})\sigma_{q}^{*}(\vec{k})i\kappa_{\alpha}\mathrm{sign}\left(z-z'\right)\mathrm{e}^{i\vec{k}\cdot\left(\vec{\rho}-\vec{\rho}'\right)}\frac{1}{2}\mathrm{e}^{-k|z-z'|}\\
    \label{eq:N_REAL_SPACE_ZZ}
    N_{z z}(\vec{r}_{p}-\vec{r}_{q})&=\frac{1}{\tilde{w}t_{p}}\int_{-\infty}^{+\infty}\frac{d^{2}\vec{k}}{(2\pi)^{2}}\int dzdz'\sigma_{p}(\vec{k})\sigma_{q}^{*}(\vec{k})k\left(\frac{1}{k}\delta(z-z')-\frac{1}{2}\right)\mathrm{e}^{-k|z-z'|}\mathrm{e}^{i\vec{k}\cdot\left(\vec{\rho}-\vec{\rho}'\right)}
\end{align}
The next step is to evaluate the two integrals:
\begin{equation}
    \int dzdz'\frac{1}{2}\mathrm{e}^{-k|z-z'|}\quad\quad \int dzdz'\frac{1}{2}\mathrm{sign}(z-z')\mathrm{e}^{-k|z-z'|},
\end{equation}
and to do that it is necessary to define the domain boundaries to $z$ and $z'$:
\begin{itemize}
    \item $z,z'\in [0,t_{p}]$ or $z,z'\in [t_{q}+\delta,2t_{q}+\delta]$: this is the case of a single layer contribution
    \item  $z\in [0,t_{p}], z'\in [t_{q}+\delta,2t_{q}+\delta]$: this corresponds to the mutual dipolar field contribution
\end{itemize}
In the first case:
\begin{equation}
\begin{aligned}
    \int_{0}^{t_{p}}dz\int_{0}^{t_{p}}\frac{1}{2}\mathrm{e}^{-k\left|z-z'\right|}dz'&=\int_{0}^{t_{p}}dz\int_{0}^{z}\mathrm{e}^{-k\left(z-z'\right)}dz'=\\
    &= \frac{1}{k}\int_{0}^{t_{p}}\mathrm{e}^{-kz}\left(\mathrm{e}^{kz}-1\right)dz=\\
    &=\frac{1}{k}\left[t_{p}-\frac{1}{k}\left(1-\mathrm{e}^{-kt_{p}}\right)\right]=\\
    &=\frac{t_{p}}{k}\underbrace{\left(1-\frac{1-\mathrm{e}^{-kt_{p}}}{kt_{p}}\right)}_{f(kt_{p})},
\end{aligned}
\end{equation}
while the second integral is equal to 0 for symmetry properties of the function.
In the second case:
\begin{equation}
    \begin{aligned}
        \frac{1}{2}\int_{0}^{t_{p}}dz\int_{t_{p}+\delta}^{t_{p}+t_{q}+\delta}dz' \mathrm{e}^{-k|z-z'|}&=\frac{1}{2}\int_{0}^{t_{p}}\mathrm{e}^{kz}dz\int_{t_{p}+\delta}^{t_{p}+t_{q}+\delta}dz'\mathrm{e}^{-kz'}=\\
        & \frac{1}{2}\frac{\mathrm{e}^{kt_{p}}-1}{k}\frac{\mathrm{e}^{-k\left(t_{p}+\delta\right)}-\mathrm{e}^{-k\left(t_{p}+t_{q}+\delta\right)}}{k}=\\
        &=\frac{1}{2}\frac{\mathrm{e}^{kt_{p}}\left(1-\mathrm{e}^{-kt_{p}}\right)\mathrm{e}^{-kt_{p}}\left(1-\mathrm{e}^{-kt_{q}}\right)}{k^{2}}\mathrm{e}^{-k\delta}=\\
        &=\frac{t_{p}}{k}\underbrace{\frac{\left(1-\mathrm{e}^{-kt_{p}}\right)\left(1-\mathrm{e}^{-kt_{q}}\right)}{2kt_{p}}\mathrm{e}^{-k\delta}}_{\tilde{f}(kt_{p})},
    \end{aligned}
\end{equation}
which in the case of $t_{p}=t_{q}=t$ yields to:
\begin{equation}
    \tilde{f}(kt)=\frac{\left(1-\mathrm{e}^{-kt}\right)^{2}}{2kt}\mathrm{e}^{-k\delta}
\end{equation}
The second integral instead will be simply equal to:
\begin{equation}
    \mathrm{sign}(z-z')\frac{t_{p}}{k}\tilde{f}(kt_{p})
\end{equation}
Now Eqs. \ref{eq:N_REAL_SPACE_XX},\ref{eq:N_REAL_SPACE_XZ},\ref{eq:N_REAL_SPACE_ZZ} can be then rewritten for cases $p=q$ and $p\ne q$:
\begin{gather}
    N_{\alpha\beta}^{(pp)}=\int_{-\infty}^{+\infty}\frac{d^{2}\vec{k}}{(2\pi)^{2}}\underbrace{\frac{\sigma_{p}(\vec{k})\sigma_{q}^{*}(\vec{k})}{\tilde{w}}\frac{\kappa_{\alpha}\kappa_{\beta}}{k^{2}}f(kt_{p})}_{N_{k,\alpha\beta}^{(pp)}}\mathrm{e}^{i\vec{k}\cdot\left(\vec{\rho}-\vec{\rho}'\right)}\\
    N_{\alpha\beta}^{(pq)}=\int_{-\infty}^{+\infty}\frac{d^{2}\vec{k}}{(2\pi)^{2}}\underbrace{\frac{\sigma_{p}(\vec{k})\sigma_{q}^{*}(\vec{k})}{\tilde{w}}\frac{\kappa_{\alpha}\kappa_{\beta}}{k^{2}}\tilde{f}(kt_{p})}_{N_{k,\alpha\beta}^{(pq)}}\mathrm{e}^{i\vec{k}\cdot\left(\vec{\rho}-\vec{\rho}'\right)}\\
    N_{\alpha z}^{(pp)}=0\\
    N_{\alpha z}^{(pq)}=\int_{-\infty}^{+\infty}\frac{d^{2}\vec{k}}{(2\pi)^{2}}\underbrace{\frac{\sigma_{p}(\vec{k})\sigma_{q}^{*}(\vec{k})}{\tilde{w}}i\frac{\kappa_{\alpha}}{k}\mathrm{sign}(z-z')\tilde{f}(kt_{p})}_{N_{k,\alpha z}^{(pq)}}\mathrm{e}^{i\vec{k}\cdot\left(\vec{\rho}-\vec{\rho}'\right)}\\
    N_{zz}^{(pp)}=\int_{-\infty}^{+\infty}\frac{d^{2}\vec{k}}{(2\pi)^{2}}\frac{\sigma_{p}(\vec{k})\sigma_{q}^{*}(\vec{k})}{\tilde{w}}(1-f(kt))\mathrm{e}^{i\vec{k}\cdot\left(\vec{\rho}-\vec{\rho}'\right)}\\
    N_{zz}^{(pq)}=\int_{-\infty}^{+\infty}\frac{d^{2}\vec{k}}{(2\pi)^{2}}\frac{\sigma_{p}(\vec{k})\sigma_{q}^{*}(\vec{k})}{\tilde{w}}(-\tilde{f}(kt))\mathrm{e}^{i\vec{k}\cdot\left(\vec{\rho}-\vec{\rho}'\right)}
\end{gather}
Considering the case of identical layers (i.e. $\sigma_{p}\sigma_{q}^{*}=|\sigma|^{2}$, the magnetometric tensor in the reciprocal space takes then the form of:
\begin{gather}
\hat{\mathbf{N}}_{k}^{(11)}=\frac{|\sigma(\vec{k})|^{2}}{\tilde{w}}
    \begin{pmatrix}
        \frac{\kappa_{x}^{2}}{k^{2}}f(kt) &\frac{\kappa_{x}\kappa_{y}}{k^{2}}f(kt)&0\\
        \frac{\kappa_{x}\kappa_{y}}{k^{2}}f(kt) &\frac{\kappa_{y}^{2}}{k^{2}}f(kt) & 0\\
        0&0&1-f(kt) 
    \end{pmatrix}\\ \hat{\mathbf{N}}_{k}^{(12)}=\frac{|\sigma(\vec{k})|^{2}}{\tilde{w}}
    \begin{pmatrix}\label{eq:TENSOR_VERT}
        \frac{\kappa_{x}^{2}}{k^{2}}\tilde{f}(kt) &\frac{\kappa_{x}\kappa_{y}}{k^{2}}\tilde{f}(kt)&i\frac{\kappa_{x}}{k}\mathrm{sign}(z_{1}-z_{2})\tilde{f}(kt)\\
        \frac{\kappa_{x}\kappa_{y}}{k^{2}}\tilde{f}(kt) &\frac{\kappa_{y}^{2}}{k^{2}}\tilde{f}(kt) & i\frac{\kappa_{y}}{k}\mathrm{sign}(z_{1}-z_{2})\tilde{f}(kt)\\
        i\frac{\kappa_{x}}{k}\mathrm{sign}(z_{1}-z_{2})\tilde{f}(kt)&i\frac{\kappa_{y}}{k}\mathrm{sign}(z_{1}-z_{2})\tilde{f}(kt)&-\tilde{f}(kt)
    \end{pmatrix}.
\end{gather}

\newpage

\subsection*{OOMMF and TetraX agreement}
\begin{figure}[h]
    \centering
    \includegraphics[scale=0.5]{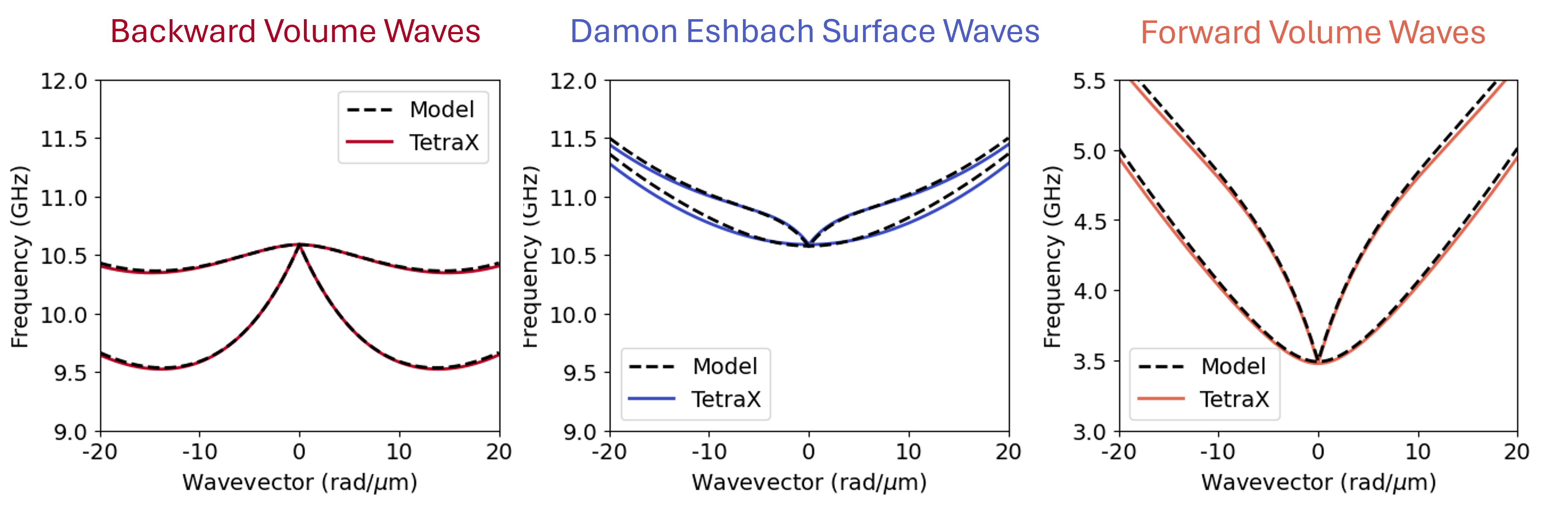}
    \caption{Agreement between the analytical model presented in the main text and TetrX micromagnetic simulations of dispersion relations for backward volume (BV), Damon-Eshbach (DE) and forward volume (FV) spin-wave configuration. The system under investigation is the same of Fig. 2 in main text, i.e. a YIG/NM/YIG hetero-structure characterized by $t=100$ nm, $\delta=4$ nm and amplitude of applied external field $H=300$ mT.}
    \label{fig:TETRAX_MODEL_AGREEMENT}
\end{figure}

\begin{figure}[h]
    \centering
    \includegraphics[scale=0.5]{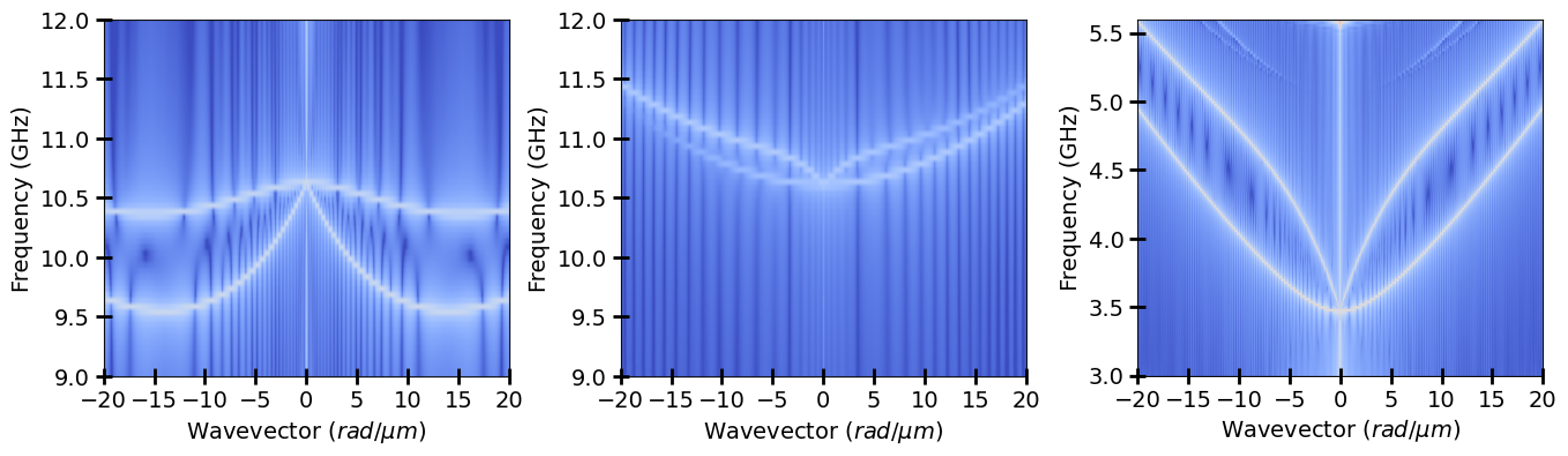}
    \caption{OOMMF micromagnetic simulations of dispersion relations for backward volume (BV), Damon-Eshbach (DE) and forward volume (FV) spin-wave configuration presented in Fig 2. of the main text, without the overlay with the analytical model. The system under investigation is the same of Fig. 2 in main text, i.e. a YIG/NM/YIG hetero-structure characterized by $t=100$ nm, $\delta=4$ nm and amplitude of applied external field $H=300$ mT.}
    \label{fig:TETRAX_MODEL_AGREEMENT}
\end{figure}

\newpage
\section*{Eigenfrequencies Derivation}\label{sec:3}
Here in the following the linear system of Eq. 9 in main text is reported in the three cases under consideration (BV,DE, FV):
\begin{equation}
    \underbrace{\begin{pmatrix}
    0& 0 & 0 & 0 & 0 & 0\\0 & 0& - \Omega_{zz}^{(11)} & - \omega_{M}F_{xz}^{(12)}  & 0 & - \omega_{M}F_{zz}^{(12)} \\\omega_{M} F_{xy}^{(11)}  & \Omega_{yy}^{(11)} & 0& \omega_{M}F_{xy}^{(12)}  & \omega_{M} F_{yy}^{(12)}  & 0\\0 & 0 & 0 & 0 & 0 & 0\\- \omega_{M}F_{xz}^{(21)}  & 0 & - \omega_{M}F_{zz}^{(21)}  & 0 & 0 & - \Omega_{zz}^{(22)}\\ \omega_{M}F_{xy}^{(21)} & \omega_{M}F_{yy}^{(21)} & 0 & \omega_{M}F_{xy}^{(22)} & \Omega_{yy}^{(22)} & i\omega
    \end{pmatrix}}_{(\hat{\boldsymbol{\mu}}_{C}\hat{\boldsymbol{\Omega}}_{C})_{BV}}\begin{pmatrix}
        m_{x}^{(1)}\\m_{y}^{(1)}\\m_{z}^{(1)}\\m_{x}^{(2)}\\m_{y}^{(2)}\\m_{z}^{(2)}
    \end{pmatrix}=-i\omega\hat{\mathbf{I}}\begin{pmatrix}
        m_{x}^{(1)}\\m_{y}^{(1)}\\m_{z}^{(1)}\\m_{x}^{(2)}\\m_{y}^{(2)}\\m_{z}^{(2)}
    \end{pmatrix}
\end{equation}
\begin{equation}
    \underbrace{\begin{pmatrix}
        0& 0 & \Omega_{zz}^{(11)} & \omega_{M}F_{xz}^{(12)} & 0 & \omega_{M}F_{zz}^{(12)} \\0 & 0 & 0 & 0 & 0 & 0\\- \Omega_{xx}^{(11)} & - \omega_{M}F_{xy}^{(11)} & 0& - \omega_{M}F_{xx}^{(12)}& - \omega_{M}F_{xy}^{(12)} & -\omega_{M} F_{xz}^{(12)} \\ \omega_{M}F_{xz}^{(21)}  & 0 & \omega_{M} F_{zz}^{(21)}  & 0& 0 & \Omega_{zz}^{(22)}\\0 & 0 & 0 & 0 & 0 & 0\\- \omega_{M}F_{xx}^{(21)}  & -\omega_{M} F_{xy}^{(21)}  & - \omega_{M}F_{xz}^{(21)} & - \Omega_{xx}^{(22)} & - \omega_{M}F_{xy}^{(22)} & 0
    \end{pmatrix}}_{(\hat{\boldsymbol{\mu}}_{C}\hat{\boldsymbol{\Omega}}_{C})_{DE}}\begin{pmatrix}
        m_{x}^{(1)}\\m_{y}^{(1)}\\m_{z}^{(1)}\\m_{x}^{(2)}\\m_{y}^{(2)}\\m_{z}^{(2)}
    \end{pmatrix}=-i\omega\hat{\mathbf{I}}\begin{pmatrix}
        m_{x}^{(1)}\\m_{y}^{(1)}\\m_{z}^{(1)}\\m_{x}^{(2)}\\m_{y}^{(2)}\\m_{z}^{(2)}
    \end{pmatrix}
\end{equation}
\begin{equation}
    \underbrace{\begin{pmatrix}
        - \omega_{M}F_{xy}^{(11)}  & - \Omega_{yy}^{(11)} & 0 & -\omega_{M} F_{xy}^{(12)}  & -\omega_{M} F_{yy}^{(12)} & 0\\\Omega_{xx}^{(11)} & \omega_{M}F_{xy}^{(11)} & 0 &\omega_{M} F_{xx}^{(12)}  & \omega_{M}F_{xy}^{(12)}  & \omega_{M}F_{xz}^{(12)} \\0 & 0 & 0 & 0 & 0 & 0\\- \omega_{M}F_{xy}^{(21)}  & - \omega_{M} F_{yy}^{(21)} & 0 & -\omega_{M} F_{xy}^{(22)}  & - \Omega_{yy}^{(22)} & 0\\\omega_{M}F_{xx}^{(21)}& \omega_{M}F_{xy}^{(21)} & \omega_{M}F_{xz}^{(21)}  & \Omega_{xx}^{(22)} &  \omega_{M}F_{xy}^{(22)} & 0\\0 & 0 & 0 & 0 & 0 & 0
    \end{pmatrix}}_{(\hat{\boldsymbol{\mu}}_{C}\hat{\boldsymbol{\Omega}}_{C})_{FV}}\begin{pmatrix}
        m_{x}^{(1)}\\m_{y}^{(1)}\\m_{z}^{(1)}\\m_{x}^{(2)}\\m_{y}^{(2)}\\m_{z}^{(2)}
    \end{pmatrix}=-i\omega\hat{\mathbf{I}}\begin{pmatrix}
        m_{x}^{(1)}\\m_{y}^{(1)}\\m_{z}^{(1)}\\m_{x}^{(2)}\\m_{y}^{(2)}\\m_{z}^{(2)}
    \end{pmatrix}
\end{equation}

Solving the eigenvalue problem yields to Eq. 11, i.e. spin-wave frequencies of the multilayer-system.
One might wonder why the coupling mechanism that occurs among layers does not result in a small linear correction of frequencies of the isolated system.
Considering the case of DE modes, and remembering that, given a block matrix $\hat{\mathbf{M}}$:
\begin{equation}
    \hat{\mathbf{M}}=\begin{pmatrix}
        \hat{\mathbf{A}} & \hat{\mathbf{B}}\\ \hat{\mathbf{C}} & \hat{\mathbf{D}}
    \end{pmatrix}
\end{equation}
it holds that:
\begin{equation}
    \mathrm{det}(\hat{\mathbf{M}})=\mathrm{det}(\hat{\mathbf{A}})\mathrm{det}(\hat{\mathbf{D}}-\hat{\mathbf{C}}\hat{\mathbf{A}}^{-1}\hat{\mathbf{B}}),
\end{equation}
where $\mathrm{det}(\hat{\mathbf{D}}-\hat{\mathbf{C}}\hat{\mathbf{A}}^{-1}\hat{\mathbf{B}})$ represents the Schur complement. The determinant of $(\hat{\boldsymbol{\mu}}_{C}\hat{\boldsymbol{\Omega}}_{C})_{DE}$ can be then written as:
\begin{equation}
    \mathrm{det}(\hat{\boldsymbol{\mu}}_{C}\hat{\boldsymbol{\Omega}}_{C}+i\omega\hat{\mathbf{I}})=\left(-\omega^{2}+\Omega_{zz}^{11}\Omega_{xx}^{11}\right)\mathrm{det}\left[\hat{\mathbf{D}}-\frac{1}{\Omega_{xx}^{11}\Omega_{zz}^{11}-\omega^{2}}\hat{\mathbf{C}}\begin{pmatrix}
        -\omega^{2} & 0 &-i\omega\Omega_{xx}^{11}\\ \omega_{M}F_{xy}^{11}\Omega_{zz}^{11}& -\omega^{2}+\Omega_{xx}^{11}\Omega_{zz}^{11} & i\omega\omega_{M}F_{xy}^{11}\\
        -i\omega\Omega_{zz}^{11} &0&-\omega^{2}
    \end{pmatrix}\hat{\mathbf{B}}\right]
\end{equation}
which clearly shows a non-linear behavior with respect to $\omega$ in the Schur complement. This evidence proves mathematically that coupling phenomena, aside from perturbation theory approaches (like also reported in main text), cannot result in a linear correction of eigenfrequencies of the system.
Even though energies are linear quantities, the dynamics (and then the torque resulting from the action of the effective field on the magnetization), together with coupling between layers, hybridize modes nonlinearly.
\newpage
\subsection*{Non trivial Coupling Description: $\Omega '$ \& $\Delta$ Frequency Corrections}
\begin{figure}[h]
    \centering
    \includegraphics[scale=0.5]{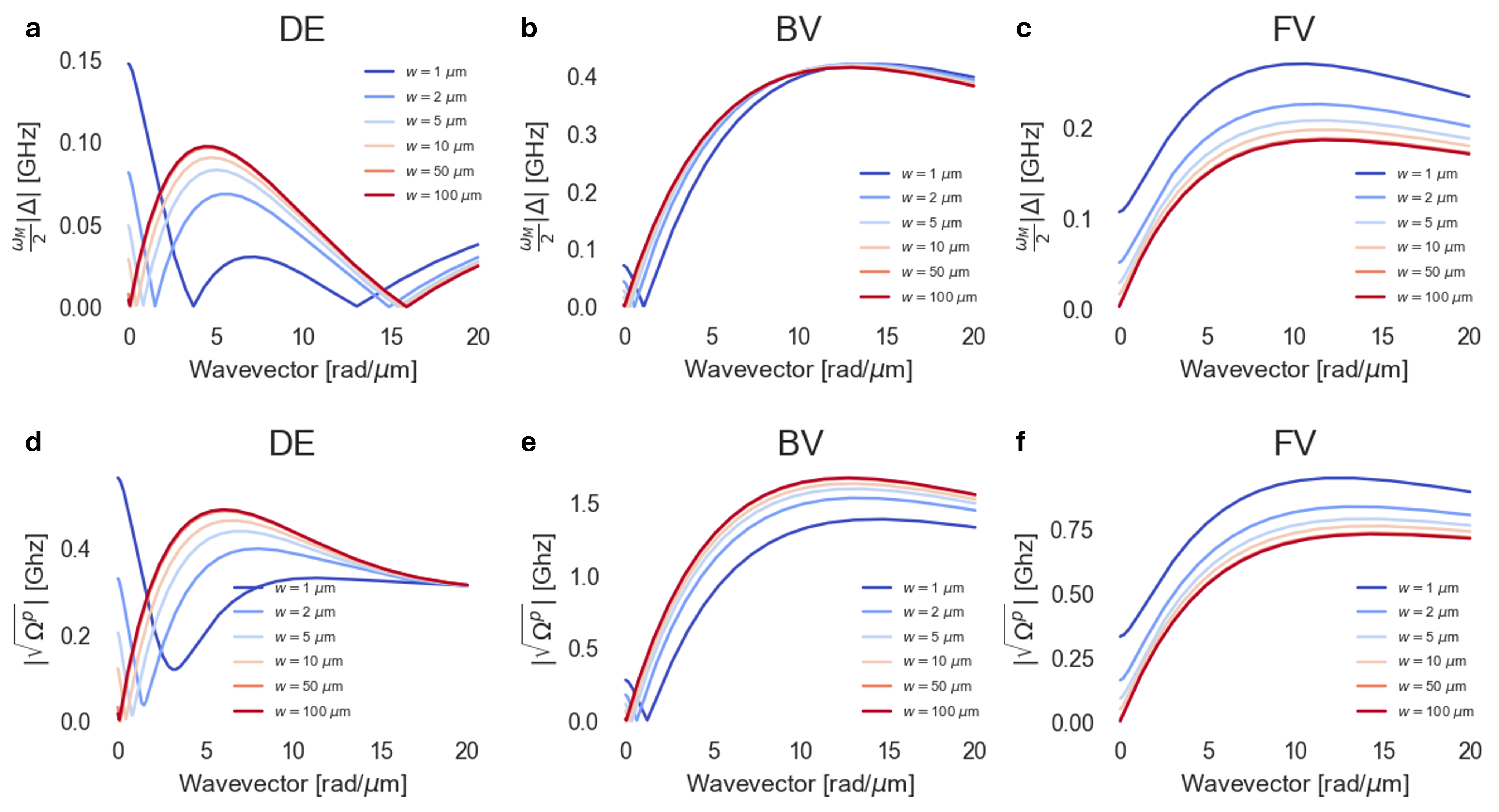}
    \caption{\label{fig:DegLIFTd}\textbf{a,b,c} Frequency correction, $\frac{\omega_{M}}{2}|\Delta|$, calculated  from the simplified perturbative model, for different widths and wavectors.\textbf{d,e,f} Evaluated $\Omega'$ as a function of the wavevector for a YIG (100~nm)/NM (4~nm)/YIG (100~nm) hetero-structure for different waveguide widths. Insets \textbf{i–iii} show the behaviour for each spin-wave configuration separately. In the Damon–Eshbach and backward-volume configurations (panels \textbf{i} and \textbf{ii}), the degenerate point is progressively shifted towards larger wavevectors as the waveguide width is reduced. For sufficiently narrow waveguides ($w=1~\mu$m), a lifting of the degeneracy is observed in the DE configuration. In contrast, in the forward-volume configuration (panel \textbf{iii}), apart from the degenerate point at $k=0$, the two branches do not intersect and no finite-$k$ degeneracy occurs}
\end{figure}
In Fig. \ref{fig:DegLIFTd} the trend in function of  $k$ for different hetero-structure widths, for DE, BV and FV configurations, is reported. In this plot, the degenerate point corresponds to the condition $\Omega\,'=0$, i.e. where the optic and acoustic branches are not separated.
An intriguing feature emerges for decreasing widths: not only does the degeneracy point shift, but it is eventually lifted. Specifically, for sufficiently small widths, the acoustic and optical branches no longer cross, but instead exhibit a minimum separation, as explicit in Fig. 5(f) in main text. This indicates that for systems no longer accountable with the thin-film approximation, the hetero-structure does not exhibit an uncoupled regime, and the dynamics of the two layers remain intrinsically connected. Another notable aspect is that, in the BV configuration, this behavior can be directly understood within the perturbative framework. Owing to the absence of off-diagonal nonreciprocal terms, the spin-wave branch of an equivalent single layer coincides with the central frequency between the acoustic and optical modes.  As a result, a Bloch-type description provides a good approximation for this configuration. 

\newpage
\subsection*{Model Extension for Lateral Dipolarly Coupled Waveguides}
\begin{figure}[h]
    \centering
    \includegraphics[width=0.5\linewidth]{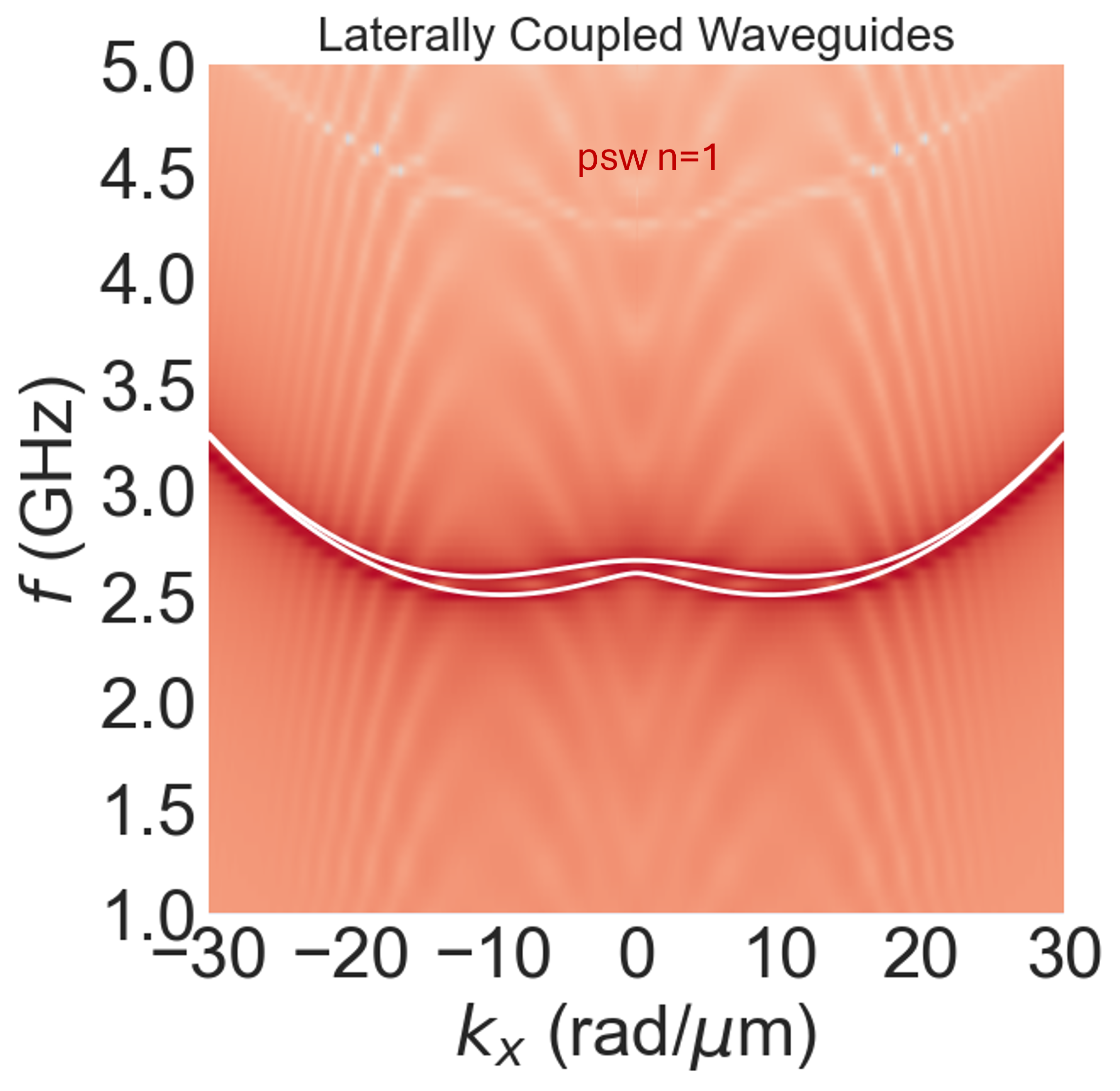}
    \caption{\label{fig:LAT}Dispersion relations of  laterally coupled waveguides in backward configuration. The system reported consists of two YIG waveguides (parameters in table IV) 100 nm wide and 50 nm thick, under and external bias field of 10 mT .}
\end{figure}

In fig. \ref{fig:LAT} the agreement between the analytical dispersion relation and the simulated one\cite{beg2022} is reported.
It is worth observing that the analytical linear system reported in \ref{sec:3} is not affected by the different system under consideration.
Main changes are associated to the mathematical representation of the magnetometric tensor. In particular, Eq. \ref{eq:TENSOR_VERT} will become $\hat{\mathbf{N}}_{k}^{(11)}=\hat{\mathbf{N}}_{k}^{(12)}$, since now $z,z'\in [0,t_{p}]$ and $\mathrm{sign}(z_{1}-z_{1})=0$.

\noindent Moreover, once the magnetometric tensor is retrieved, the $k_{y}$ dependence is removed by anti-transforming along $y$-direction and evaluating at fixed $y=y_{p}-y_{q}$:
\begin{equation}
    \hat{\mathbf{F}}_{k_{x}}^{(pq)}=\frac{1}{2\pi}\int_{-\infty}^{+\infty}\hat{\mathbf{N}}_{k}^{(pq)}\mathrm{e}^{ik_{y}(y_{p}-y_{q})}dk_{y},
\end{equation}
as reported in main text through Eq. 5 in the case of vertically aligned magnetic layers (i.e. $y_{1}=y_{2}$).

\newpage
\subsection*{Forward Volume Modes: Geometrical Analysis}
\begin{figure}[h]
   \centering
    \includegraphics[width=\linewidth]{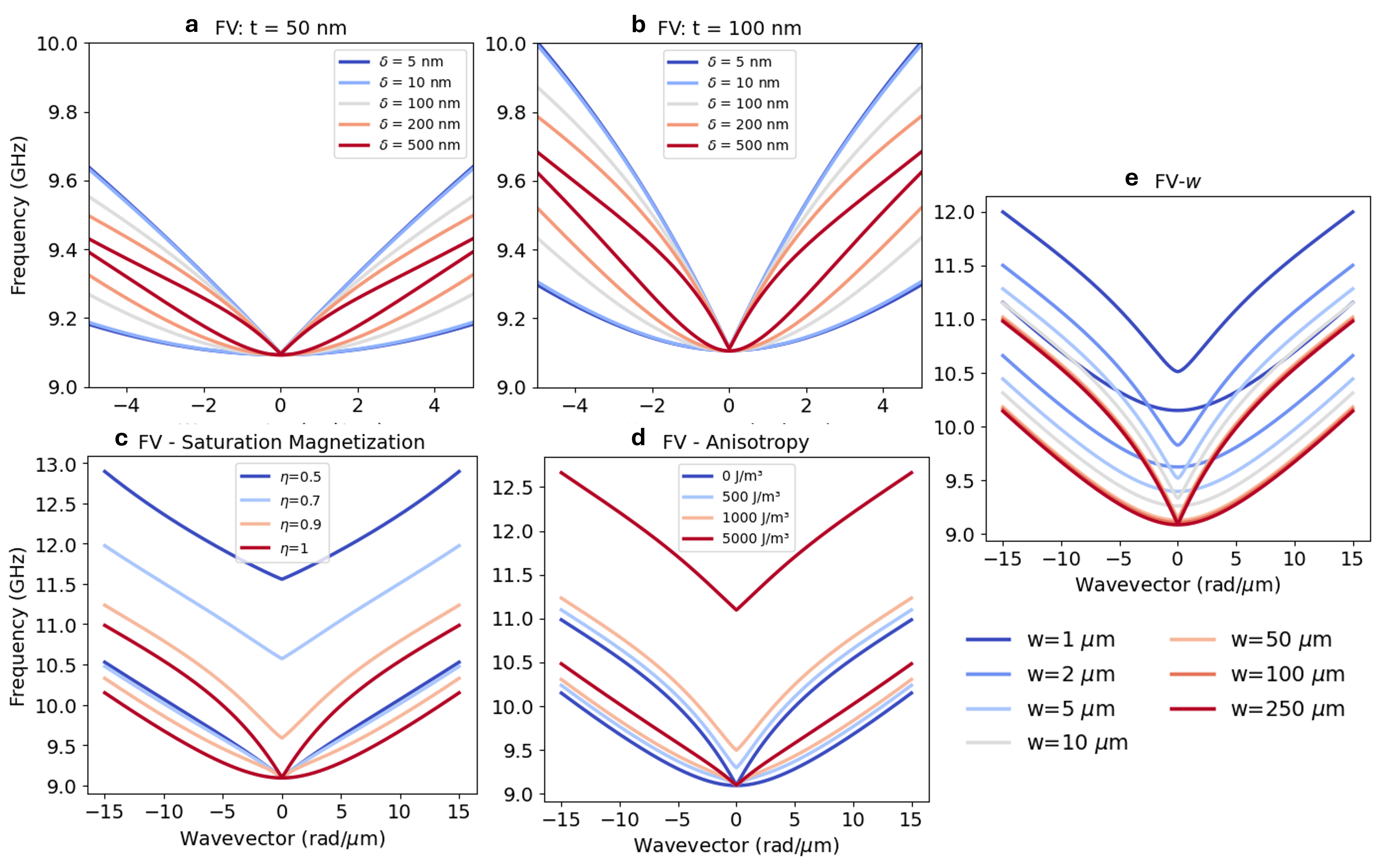}
    \caption{\textbf{a},\textbf{b} Calculated spin-wave dispersion relations in FV configuration for two identical infinite YIG layers with thickness t = 50, 100 nm (parameters in Table IV in main text), separated by a non-magnetic spacer of increasing thickness $\delta$. 
    \textbf{c} Spin-wave dispersion relations in FV configuration for two YIG layers with different saturation magnetizations, where $M_{s,p}=\eta M_{s,q}$ and $\eta\leq1$, in the absence of magnetic anisotropy. 
    \textbf{d} Spin-wave dispersion relations in the forward volume configurations for two YIG layers with identical saturation magnetization, where the top layer hosts a perpendicular magnetic anisotropy modeled as a uniaxial anisotropy along $u = (0, 0, 1)$. Results are shown for different values of the anisotropy constant $K$. 
    \textbf{e} Calculated spin-wave dispersion relations of a YIG (100 nm)/NM(4 nm)/YIG (100 nm) hetero-structure, with parameters reported in Table IV, as a function of the waveguide width.}
    \label{fig:FW_Global}
\end{figure}

In Fig. \ref{fig:FW_Global} the studies presented in the main text are also reported for the forward volume configuration. Here, it is considered the case of an infinite film under an external field of 500 mT. Similar conclusions with DE and BV configurations can be drawn. Increasing the spacer thickness progressively reduces the separation between the acoustic and optic branches, approaching the single-waveguide limit, consistent with the decay of dynamic dipolar coupling with interlayer distance (panel a,b). As the mismatch between the saturation magnetizations of the two layers increases, the separation between the acoustic and optic branches increases, with the acoustic branch shifting to lower frequencies (panel c).
The introduction of anisotropy produces an effective asymmetry between the layers, leading to an increased separation between the branches, reproducing the same qualitative trend observed (panel d).
Consistent with the FV configuration, the increasing demagnetizing contribution shifts the dispersion branches upward for narrower waveguides. As explained in the main text the FV configuration doesn't present a degenerate point at which the symmetric and antisymmetric wavevectors coincide (panel e).

\end{document}